\documentclass[aps,prb,twocolumn,superscriptaddress]{revtex4-2}
\usepackage{graphicx}
\usepackage{mathptmx}
\usepackage{amsmath}
\usepackage[]{graphicx}
\usepackage{bm}
\usepackage{times}
\usepackage{tabularx}
\usepackage{color}
\usepackage{dcolumn}
\usepackage{amsmath}
\usepackage{amssymb}
\usepackage{amsfonts}
\usepackage{calc}
\usepackage{times}
\usepackage{tabularx}
\usepackage{xcolor,colortbl}
\usepackage{rotating}
\usepackage[colorlinks,citecolor=blue,linkcolor=blue]{hyperref}
\usepackage{amsmath}
\usepackage{hyperref} 
\usepackage{soul}
\usepackage{caption}
\usepackage{subcaption}
\usepackage{upgreek}
\usepackage[normalem]{ulem}
\begin{document}

\title{Near complete laser-induced modulation of the ferromagnetic-antiferromagnetic phase fraction in FeRh films}
\author{Alexis Pecheux}
\email{Contact author: alexis.pecheux@ens-paris-saclay.fr}
\affiliation{Université Paris-Saclay, ENS Paris-Saclay, CNRS, SATIE, 91190 Gif-sur-Yvette, France }
\author{Robin Salvatore}
\affiliation{
Institut des Nanosciences de Paris, Sorbonne-Universit\'{e}, CNRS , 4 Place Jussieu, F75005 Paris, France}
\author{Laura Thevenard}
\affiliation{
Institut des Nanosciences de Paris, Sorbonne-Universit\'{e}, CNRS , 4 Place Jussieu, F75005 Paris, France}
\author{Jon Ander Arregi}
\affiliation{CEITEC BUT, Brno University of Technology, Purkyňova 123, 612 00 Brno, Czech Republic}
\author{Vojt\v{e}ch Uhl\'{i}\v{r}}
\affiliation{CEITEC BUT, Brno University of Technology, Purkyňova 123, 612 00 Brno, Czech Republic}
\affiliation{Institute of Physical Engineering, Brno University of Technology, Technická 2, 61669 Brno, Czech Republic}
\author{Morgan Almanza}
\affiliation{Université Paris-Saclay, ENS Paris-Saclay, CNRS, SATIE, 91190 Gif-sur-Yvette, France }
\author{Danièle Fournier}
\affiliation{
Institut des Nanosciences de Paris, Sorbonne-Universit\'{e}, CNRS , 4 Place Jussieu, F75005 Paris, France}
\author{Catherine Gourdon}
\affiliation{
Institut des Nanosciences de Paris, Sorbonne-Universit\'{e}, CNRS , 4 Place Jussieu, F75005 Paris, France}
\author{Martino LoBue}
\affiliation{Université Paris-Saclay, ENS Paris-Saclay, CNRS, SATIE, 91190 Gif-sur-Yvette, France }

\date{\today}

\begin{abstract}
With its huge entropy change and a strong interplay between magnetic order, structural and electrical properties, the first-order antiferromagnetic/ferromagnetic phase transition is a paradigmatic example of the multicaloric effect. The unraveling of the physics underlying the phase transition needs a better understanding of the thermal hysteresis of FeRh within the AF-FM phase coexistence region. In this work, we compare the effect of two very different types of thermal cycling on the hysteresis of the magnetic order: quasi-static heating, and cooling of the entire $195$~nm thick film, and a $f$=100~kHz  modulated heating driven by a laser focused down to a spot of about ten microns squared at the film surface. Taking advantage of the reflectivity difference between both phases to probe optically their respective fraction, we show that whereas only temperature-driven reflectivity variations ($dR/dT$, thermoreflectance)  are detected in the pure phases, a huge modulation of the phase-dependent reflectance at the driving frequency $f$ is detected in the phase coexistence temperature range. This is quantitatively described as resulting from a substantial modulation of the FM fraction (up to 90$\%$) with increasing laser power. A simplified rate-independent hysteresis model with return-point-memory (RPM), represented in terms of bistable units that undergo a temperature excursion corresponding to a given laser power, reproduces very well the optically measured FM phase modulation characteristics for a broad range of temperature excursions. This offers an insight into the leading role of quenched disorder in defining thermal hysteresis in FeRh under high excitation frequency, when the material is periodically driven out-of-equilibrium.
\end{abstract}

\maketitle

\section{Introduction}

Almost $90$ years since the first report of its ferromagnetic ordering on heating \cite{Fallot1938, Fallot1939-1}, nearly equiatomic FeRh alloy still arouses a considerable interest in understanding its phenomenology, and harnessing its multifunctional properties towards new applications \cite{Lewis2016-1}. The first-order phase transition of FeRh from antiferromagnetic (AF) to ferromagnetic (FM) order when the temperature rises above $\approx 370$~K, shows a rather wide thermal hysteresis $\approx 10$~K, and is accompanied by a $1~\%$ volume change associated with an iso-structural lattice expansion. Besides, the strong interplay between magnetic, electronic and structural degrees of freedom \cite{Lewis2016-1}, offers unique opportunities for probing, and controlling the phase transition through different external stimuli. Indeed, the huge entropy change (\textit{i.e.} up to $12.5$~J\thinspace kg$^{-1}$K$^{-1}$ \cite{Lyubina_2017}) associated with the phase transition makes FeRh a paradigmatic example of the so-called giant multicaloric materials sought after to be used as a refrigerant in solid-state cooling devices \cite{Stern-Taulats2017-1}. 

In addition, the persistence of an abrupt phase change in the thin film form offers the opportunity to investigate the transition kinetics, and the phase coexistence at a scale close to the size of the phase-domains \cite{Maat2005-1, Uhlir2016, deVries2014-1}, with particular attention devoted to the coupling between magnetic order and strain \cite{Arregi2020a, Cherifi2014-1, Loving2017-1}. Furthermore, thin epitaxial FeRh films have sparked a renewed interest for potential applications such as thermally assisted magnetic recording \cite{Thiele2003-1}, memory cells \cite{Marti2014-1}, and microscale patterning \cite{Mei2020a}. 

Beside the lattice parameter \cite{deVries2014-1, Arregi2020a}, and the electrical resistivity \cite{Kouvel1962-1, Uhlir2016, Wu2024-1}, the interplay between FeRh optical properties and magnetic ordering represents an additional tool for probing hysteresis, and phase coexistence across the phase transition. More precisely, the $\approx 3 - 5 \%$ difference between the reflectivity of the FM and  AF phases \cite{Saidl2016a} has been used to spatially resolve the phase change showing a rather broad distribution of transition temperatures, at the micrometer scale, presumably due to the presence of structural defects, and to the compositional hypersensitivity of FeRh \cite{Staunton2014-1, Aubert-2024}. The same property has been used, through high resolution optical microscopy, for capturing unprecedented insights into the phase domain patterns, and growth, as well as into their dependence on temperature and field \cite{Arregi2023-1}. FeRh films have also been used as a reference system for investigating laser-induced ultrafast phase transformations \cite{Thiele2004b, Ju2004, Bergman2006, Radu2010, Quirin2012a, Mariager2012-1, Pressacco2018,Pressacco2021, Mattern2024a,Dolgikh2025}. In particular, picosecond time-domain TR measurements have recently been used to investigate the non-equilibrium interplay of electronic, lattice, and magnetic degrees of freedom close to the phase transition\cite{Harton2024}. 

Modulated thermoreflectance experiments have recently served for measuring the thermal conductivity of FeRh and its dependence on magnetic ordering \cite{Castellano2024}. Here, through the same experimental setup presented as in Ref. \cite{Fournier2020}, we show that in a standard laser-modulated reflectance experiment there is an additional contribution to the usual thermoreflectance signal. This additional contribution, up to $60$ times larger than the thermoreflectance, is present only in the AF-FM phase coexistence temperature interval and is strongly dependent upon the power of the pump laser. Since the pump laser is a modulated continuous-wave laser, we show through suitable thermal modeling that its main effect is to induce a time-dependent periodic temperature change in the material over the thermal diffusion length. Indeed, one of the key findings of this work is to show that the origin of the additional contribution to the signal is the modulation of the FM phase fraction, taking place within the phase coexistence interval, driven by the laser-induced time-dependent temperature change.

This phase-modulated reflectance signal is then used to study the FM-AF first-order phase transition kinetics, and its thermal hysteresis at the scale of the laser spot. Using a thermal model jointly with the thermal properties reported in Ref.\cite{Castellano2024}, the time dependence of the temperature over the spot, controlled by the pump laser, is calculated and related to the phase-modulated signal. In this way thermal hysteresis is studied under a periodic temperature driving with heating/cooling rates that would be hardly achievable using standard thermal controllers. The resulting data are used to understand to what extent the key thermal hysteresis properties measured over the whole sample, and in thermal equilibrium with a heating stage, are still verified at the local scale, with the probed material driven out of thermal equilibrium by fast temperature pulses (\textit{i.e.} several tens of degrees in less than a microsecond). 

The paper is organized as follows. In the next section, \ref{sec:Setup}, the sample preparation and experimental setup are described. In section \ref{sec:MR} the temperature-dependent reflectance data are presented with particular attention to the thermal hysteresis exhibited by the reflectance. The relationship between reflectance, and phase fraction across the transition is discussed.  In section \ref{sec:model} a simplified rate-independent hysteresis model showing return-point-memory (RPM) is used to qualitatively reproduce the phenomenology presented in the preceding sections. The last section, \ref{sec:Conclusion}, is devoted to a general discussion on the main findings presented in this work, and on the perspectives they are opening.

\section{\label{sec:Setup}Sample and experimental setup\protect}

\begin{figure}
\centering
		\includegraphics[width=0.98\columnwidth]{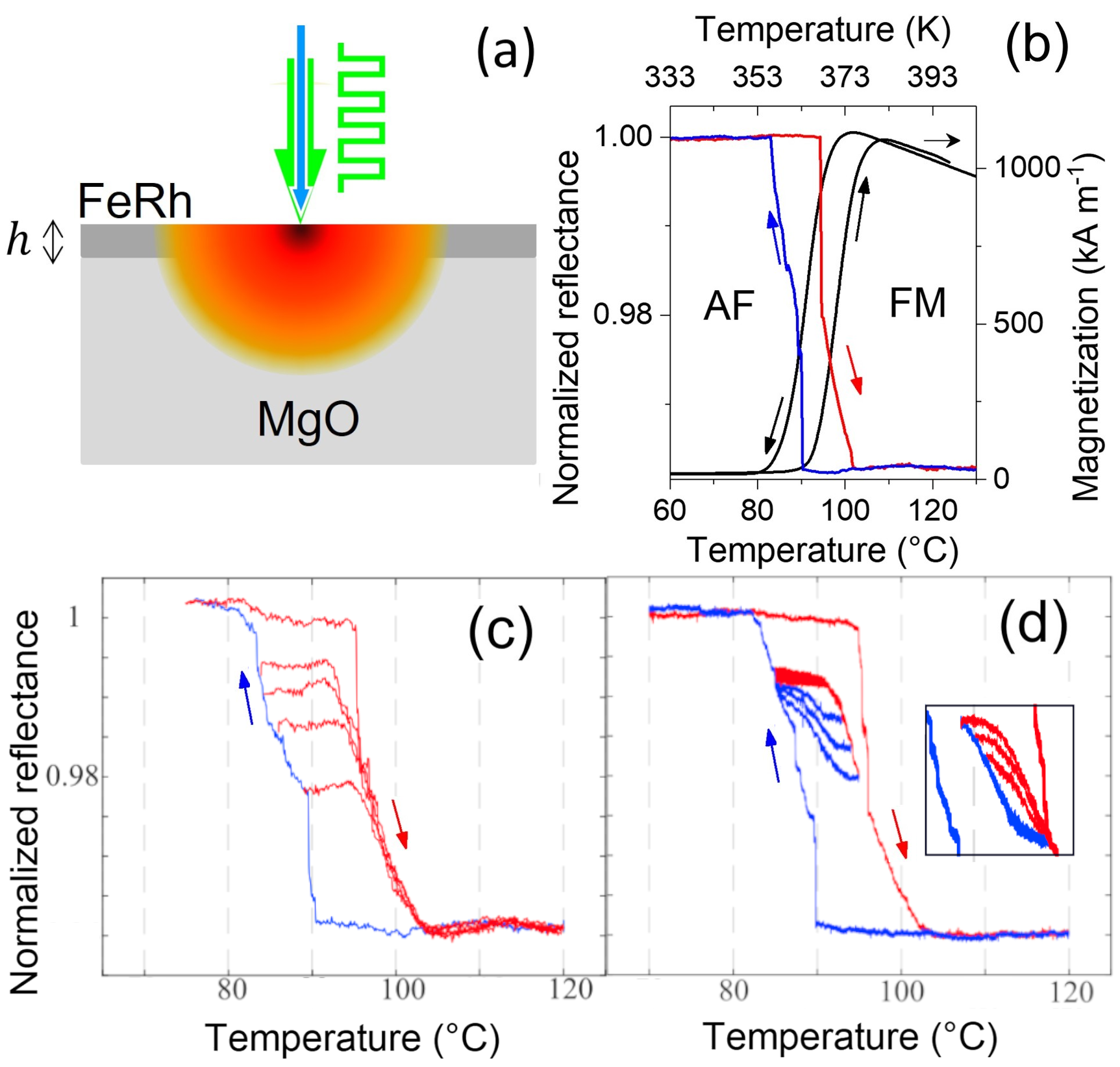}
\caption{\label{fig:Fig1}(a) Scheme of the modulated reflectance experiment. (b) Magnetization (black curve) and reflectance of the FeRh/MgO sample normalized to the AF phase level at  $\lambda_b = 488$~nm  on an area of diameter $4~\upmu$m (smoothed signal), acquired from the ADC channel of the lock-in amplifier, with pump laser off, along a heating (red curve), and cooling (blue curve) ramp driven by the heating stage within the quasi-static regime. (c) Quasi-static major hysteresis loop with a set of first-order return curves measured without pump laser. The return curves are measured over the cooling branch starting from $T_{start} = 120$\textdegree C, and heating back to $T_{start}$ at a reversal point $T_{rev} = 89, 86, 85, 84$\textdegree C; (d) a set of nested minor loops starting from the cooling branch verifying the return point memory (RPM). The inset shows a set of nested minor loops starting from the heating branch. A detailed description of the protocol followed for measuring the nested minor loops is given in the main text.}
\end{figure}

All the measurements have been carried out on a FeRh sample prepared by magnetron sputtering of an equiatomic FeRh target, grown epitaxially on a MgO (001) substrate. The film, of thickness $h=195$~nm, is protected from oxidation by a $2$~nm Pt cap layer. Details on the sample growth can be found in Ref. \cite{Castellano2024}. From this process, a (001) textured FeRh film is obtained \cite{Arregi2020a}, and stoichiometry across the whole thickness has been confirmed by STEM-EDX mapping \cite{Arregi2023-1}. The magnetization hysteresis cycle shown in Fig.~\ref{fig:Fig1}(b) has been obtained using a vibrating sample magnetometer.

All the modulated reflectance experiments have been performed with the microscopy setup presented in Ref. \cite{Fournier2020}. The sample is placed in a Linkam THMS600 heating stage set under a microscope. The reflectance of the sample is monitored using a probe laser beam (wavelength $\lambda_b = 488$~nm) focused onto the sample surface using a microscope objective with numerical aperture $NA=0.3$. The pump laser (wavelength $\lambda_g=532$~nm) is focused onto the sample using the same microscope objective (Fig.\ref{fig:Fig1} (a)). The pump and probe spots diameters have been estimated to be 4 $\upmu$m. The polarization scheme of the set-up~\footnote{Both pump and probe beams are circularly polarized when impinging the sample, and linearized back when returning towards the photodiode. A narrow interference filter in front of the detector suppresses the green pump beam.} makes the signal insensitive to either in-plane or out-of-plane components of the magnetization. The amplitude of the pump laser is square-modulated using an acousto-optic modulator at frequency $f = 100$~kHz. The pump power is varied between $0.29$~mW and $22.6$~mW. The probe power is kept fixed at $0.17$~mW. 
The reflected probe laser flux having a continuous (DC) component and a modulated one at the frequency $f$ of the pump laser is detected using a New Focus 1801 photodiode and acquired by a SR7280 lock-in amplifier. The recorded in-phase $V_{\rm{f}}^{\rm{p}}$ and in-quadrature $V_{\rm{f}}^{\rm{q}}$ voltages are used to obtain the amplitude $V_{\rm{f}}=\sqrt{\left(V_{\rm{f}}^{\rm{p}}\right)^2+\left(V_{\rm{f}}^{\rm{q}}\right)^2}$ and the phase $\phi=\arctan\left(V_{\rm{f}}^{\rm{q}}/V_{\rm{f}}^{\rm{p}}\right)$ ~\footnote{A filter cutting the green laser is placed in front of the photodiode. Prior to the experiment, the phase of the lock-in detection is adjusted by removing the filter and detecting directly the reflected pump laser flux with the photodiode.}. The pump and probe overlap is adjusted by scanning the probe spot across the pump spot in order to maximize the modulated reflectance voltage $V_{\rm{f}}^{\rm{p}}$ in the pure AF or FM phase. The DC reflectance voltage $V_0$ is monitored either using the ADC input of the lock-in amplifier or by processing the photodiode signal with a NI USB-6281 acquisition card using an anti-aliasing filter ($44$~kHz),
both methods giving identical hysteresis curves. 
No differences between the reflectance curves have been observed for temperature scan rates $\leq 50$~\textdegree C/min. All the temperature changes driven by the heating stage have been performed at rates $dT/dt \leq 10$~\textdegree C/min, namely, well within a velocity regime that can be considered as quasi-static. No magnetic field is applied when doing the reflectance measurements.

\section{\label{sec:MR} Measurements \protect}
\begin{figure*}[ht]
\includegraphics[width=1\textwidth]{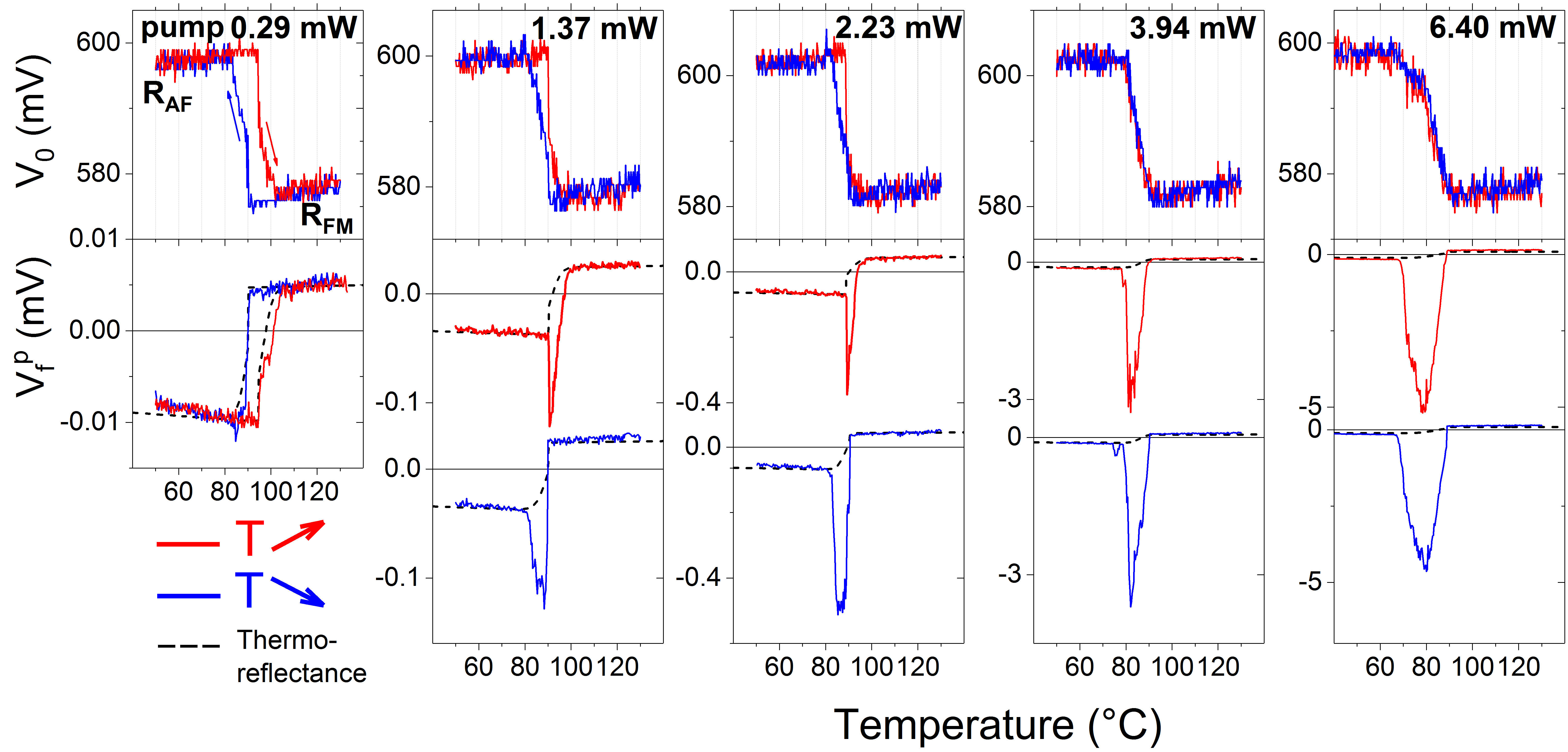}
\caption{\label{fig:DC-AC-R}(a) to (e): Reflectance thermal hysteresis curves acquired from the lock-in ADC channel as a function of heating stage temperature for increasing pump power. A progressive closing of the cycle and global down-shift in temperature are apparent. (f) to (n): Modulated reflectance in-phase signal  
for up (down) temperature ramp in red (blue) colour. For higher pump powers the modulated reflectance is similar to (j,n) albeit wider and with the transition starting at a lower temperature. The dashed black curves represent the calculated contribution of the thermoreflectance to the signal (see text).
}
\end{figure*}

Jointly with the magnetization versus temperature curve, Fig.~\ref{fig:Fig1}(b) shows the normalized reflectance probed by the focused blue laser (without pump laser) with heating and cooling curves in red and blue, respectively. The reflectance shows a temperature hysteresis of similar width to that measured by magnetometry, albeit with sharper AF-FM and FM-AF transitions. The small but significant reflectance drop of $3.7\%$ from the AF to the FM phase, in agreement with previous measurements \cite{Saidl2016a}, makes the measured reflectance value a probe of the relative phase fraction over the spot size.   

Figures~\ref{fig:Fig1}(c,d) show more thermal hysteresis curves measured changing the temperature in quasi-static conditions, and in the absence of pump laser.  Fig.~\ref{fig:Fig1}(c) shows a set of first-order reversal curves departing from the major cooling branch (blue). The cooling starts at $T = 120$~\textdegree C, where the sample is in the pure FM state. Each reversal curve is measured by stopping the cooling at a given reversal temperature, $T_{rev} = 89, 86, 85, 84~$\textdegree C (Fig.~\ref{fig:Fig1} (c)), and heating back to $120$~\textdegree C. Similar curves have been measured starting from the major heating branch in order to verify the stability, under quasi-static conditions, of the major hysteresis cycle (not shown here). Figure~\ref{fig:Fig1}(d), and its inset, show two sets of nested minor loops starting from the cooling, and from the heating branch respectively. The first one have been measured by cooling the sample from $T = 120$~\textdegree C down to $T_{rev} = 85$~\textdegree C, and afterward performing a set of heatings up to $T_i$ followed by a cooling back to $T_{rev}$. The second one by heating the sample from $70$~\textdegree C up to $96$~\textdegree C, and subsequently measuring a set of cooling down to $T_i$ followed by a heating back to $T_{rev}$. In both cases ten minor loops have been measured keeping a one degree difference between $T_i$ and $T_{i+1}$ with $i = 1, 2 ..., 10$ (for the sake of clarity only three minor loops are plotted in the figure, and in the inset).
Similar measurements performed at different temperatures starting on both major cooling and heating branches have been performed in order to check compliance with the RPM, a key hysteresis feature that will be discussed in what follows.

\begin{figure}
\centering
		\includegraphics[width=1\columnwidth]{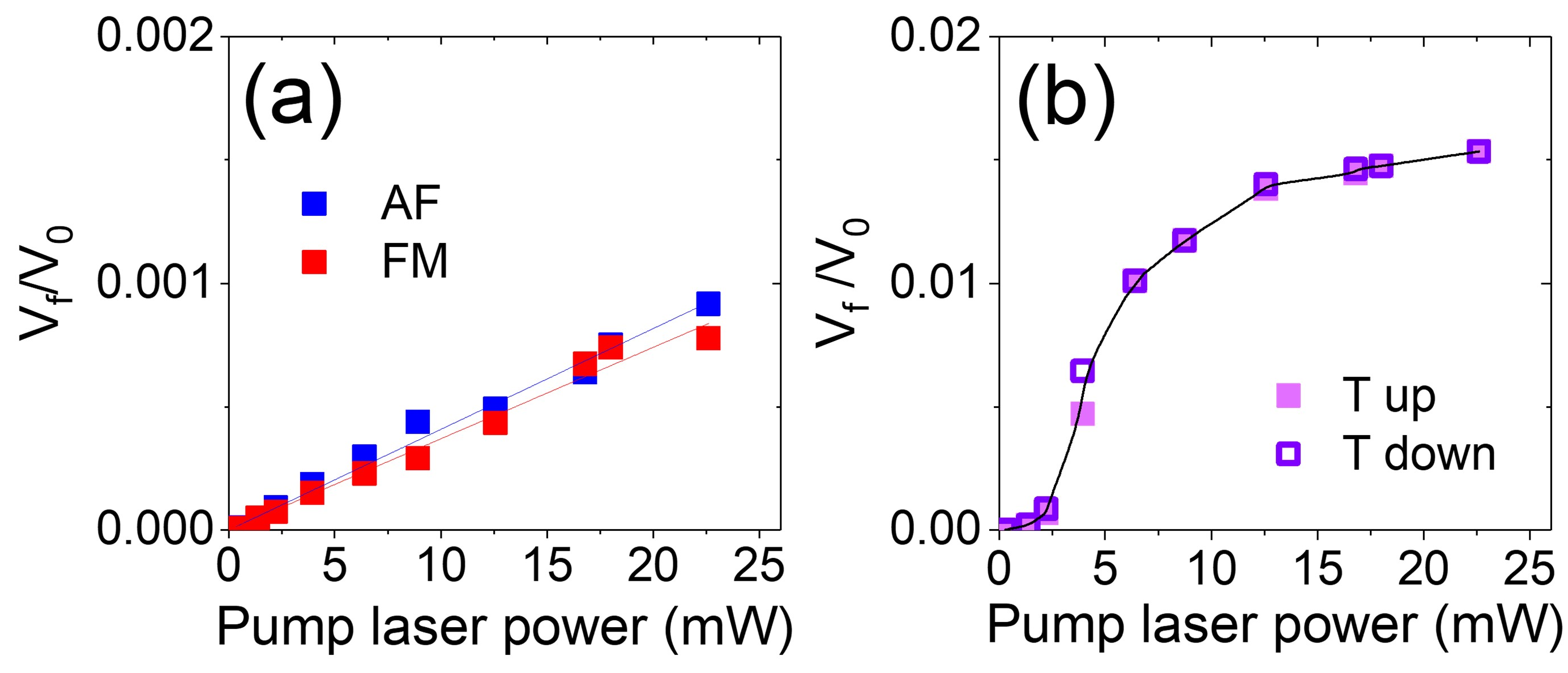}
\caption{\label{fig:lin-nonlin}Modulated reflectance amplitude for overlaying pump and probe laser spots versus
pump power $P$: (a) measured at a temperature where the sample is either in the pure AF ($T$=50$^{\circ}$C (40$^{\circ}$C) for $P<$13\thinspace mW ($>$13\thinspace mW)), or in the pure FM phase (110 $^{\circ}$C), and (b) in the phase coexistence interval, at the temperature where $V_{\rm{f}}$ is maximum (the full curve is a guide for the eye). Note the factor 10 between $y$-scales in (a) and (b).}
\end{figure}

Figure~\ref{fig:DC-AC-R} shows measurements carried out with the lock-in amplifier when the pump and probe laser spots are superimposed and the pump modulates the temperature using a $100$~kHz square-wave excitation at different pump powers. At the lowest power, $0.29$~mW (Fig.\ref{fig:DC-AC-R} (a)), the DC signal exhibits a thermal hysteresis limit cycle as a function of the heating stage temperature, similar to that measured in quasi-static conditions without modulation, as shown in Fig.\thinspace \ref{fig:Fig1}(b). On the contrary, under increasing pump power, the heating branch of the DC-signal hysteresis loop gets closer to the cooling branch till completely collapsing there at $3.94$\thinspace mW as shown in Fig.\thinspace \ref{fig:DC-AC-R}(b-e). At higher pump powers the DC reflectance is similar in shape to that of Fig.\thinspace \ref{fig:DC-AC-R}(e) albeit wider, with a transition starting at lower temperature. 
The modulated reflectance in-phase signal $V_{\rm{f}}^{\rm{p}}$ for superimposed pump and probe is shown in Fig.\thinspace \ref{fig:DC-AC-R}(f)-(n). At the lowest pump power (Fig.\thinspace\ref{fig:DC-AC-R}(f)) $V_{\rm{f}}^{\rm{p}}$ is of the order of $10~\upmu$V (\textit{i.e.} about $10^{-5}$ times the reflectance). Assuming that this signal arises from thermoreflectance with $\Delta R=(dR/dT)\Delta T$ the negative signal in the AF phase and positive signal in the FM phase means that the derivatives $dR_{AF}/dT$ and $dR_{FM}/dT$ are negative and positive, respectively. As the pump power increases, an additional contribution to $V_{\rm{f}}$ develops, boosting the signal amplitude up to a few millivolts. At temperatures where the material is in the pure AF or FM phase the normalized modulated signal $V_{\rm{f}}$ shows a linear dependence over the pump laser power up to $22.6$~mW, as shown in Fig.\thinspace \ref{fig:lin-nonlin}(a), in agreement with the thermoreflectance model where the temperature increase is proportional to the incident heat flux \cite{Fournier2020,Castellano2024}. On the contrary, in the mixed AF-FM phase temperature interval, the $V_{\rm{f}}$ signal shows a strong non-linear dependence on the pump power (Fig.~\ref{fig:lin-nonlin}(b)), hinting at an additional contribution to the thermoreflectance.

\subsection{\label{subsec:Rvsx}Reflectance and phase fraction}

Using the reflectance value as a measure of the phase fraction over the spot allows the study of the thermal hysteresis associated with the first-order transition when the temperature is quasi-statically driven by the heating stage controller, as well as when the FM and AF phases are modulated by the pump laser. In the following we shall assume that the contribution to the reflectance $R$ of the AF and FM domains within the penetration depth of the laser light can be expressed as: 
\begin{equation}
    R = \frac{1}{A}\left(R_{AF}\sum_{i} A_{i}^{AF} +R_{FM} \sum_{j} A_{j}^{FM}\right),
       \label{eq:R}
\end{equation}

where $A_{i}^{AF} (A_{j}^{FM}$) refer to a domain area in the AF (FM) phase, and $A$ is the total area probed by the laser spot. As both the number and the area of the domains of a given phase depend on $T$ \cite{Keavney2018-1}, the total FM phase fraction can be written as $x(T) = \sum_i A_i^{FM}/A$. This allows to recast Eq. (\ref{eq:R}) as:

\begin{equation}
     R(T) =  R_{AF}(T)\left(1-x(T)\right) + R_{FM}(T) x(T).
     \label{eq:R(T,x)}
\end{equation}

Therefore, for each pump power, the $x(T)$ fraction is obtained from the temperature dependent DC voltage (Fig.~\ref{fig:DC-AC-R}(a)-(e)) as $x(T)=\left(V_0(T)-V_0^{AF}\right)/\left(V_0^{FM}-V_0^{AF}\right)$, where $V_0^{AF}$ ($V_0^{FM}$) is the DC voltage in the AF (FM) phase. From Fig.\thinspace \ref{fig:DC-AC-R}(f)-(n) we see that the laser-induced variation of the reflectance $\Delta R$ presents two contributions.  One is the laser-driven temperature modulation $\Delta R^{th}$ (\textit{i.e.} the thermoreflectance contribution) \cite{Castellano2024}, depending on the derivatives  $dR_{AF}/dT$  and $dR_{FM}/dT$,  which describes quite well the data at the lowest pump power. As we will show hereafter, the other one arises from the modulation of the FM fraction $x$, $\Delta R^x$, depending on  $dR/dx=-R_{AF}+R_{FM}$, which kicks in at higher power and gives a much larger $V_{\rm{f}}$ signal in the mixed phase.

In order to single out this additional signal we must first subtract the thermoreflectance contribution $\Delta R^{th}$.

In the mixed phase Eq.(\ref{eq:R(T,x)}) allows to depict $\Delta R^{th}$ as the sum of the thermoreflectance signals of each phase, weighted by the surface fraction they represent under the laser spot. This can be written as:
\begin{equation}
    \Delta R^{th}=\frac{1}{A}\left(\frac{dR_{AF}}{dT}\sum_{i} A_{i}^{AF} \Delta T_{AF}+\frac{dR_{FM}}{dT} \sum_{j} A_{j}^{FM} \Delta T_{FM}\right),
\end{equation}
hence
\begin{equation}
    \Delta R^{th}=\frac{dR_{AF}}{dT}(1-x) \Delta T_{AF}+\frac{dR_{FM}}{dT} x \;\Delta T_{FM}\;, 
     \label{Eq:DeltaRth} 
\end{equation}
where $\Delta T_{AF}$ ($\Delta T_{FM}$) is the temperature variation in the AF (FM) phase.  

Consequently, the thermoreflectance contribution to the in-phase ($i = p$), and in-quadrature ($i = q$) signals is: $(1-x)V_{\rm{f}}^{i}(AF)+ x\; V_{\rm{f}}^{i}(FM)$, where $V_{\rm{f}}^{i}(FM)$  and $V_{\rm{f}}^{i}(AF)$ are the $f$-harmonic contributions to the thermoreflectance signal within the pure FM phase ($T$=110 $^{\circ}$C) and AF phase ($T$=50$^{\circ}$C (40$^{\circ}$C) for $P<$13\thinspace mW ($>$13\thinspace mW)), respectively. A more detailed analysis presented in Appendix~\ref{App:TemperatureIncrease} allows the determination of the thermal parameters. The derivatives $dR_{AF}/dT$ and $dR_{FM}/dT$ are rather small, of the order of$-2.8\thinspace10^{-5}$~K$^{-1}$ and $1.6 \times 10^{-5}$~K$^{-1}$, respectively. Let us note that the apparent slope of $R_{AF}$ or  $R_{FM}$ in Fig.~\ref{fig:DC-AC-R}(a)-(e) is not a true temperature dependence but results from the local variation of the reflectance amplitude and/or from slight defocusing of the laser spot due to a small shift of the sample holder under the laser spot.

The  thermoreflectance contribution $\Delta R^{th}_{\rm{f}}$ to the 
measured signal $V^{\rm{p}}_{\rm{f}}$ calculated using the $x(T)$ 
fraction determined as explained above is shown as a black dashed line in Fig.~\ref{fig:DC-AC-R}(f)-(n). Whereas $\Delta R^{th}_{\rm{f}}$ accounts for almost $100\%$ of the signal at low power, it barely represents  a few $\%$ of it at the higher ones (\textit{i.e.} above $2$~mW).

\begin{figure}[ht]

\includegraphics[width=0.9\columnwidth]{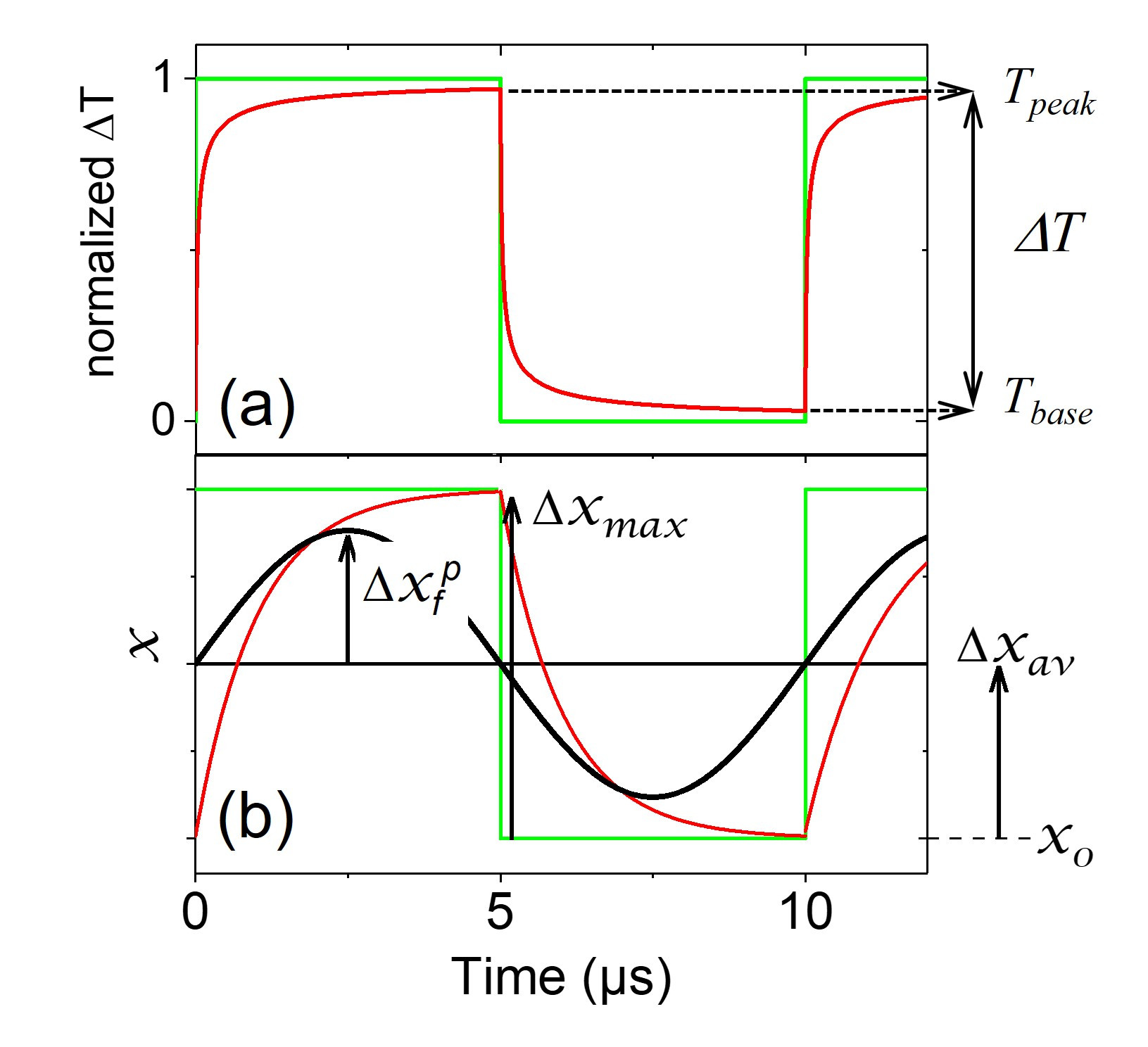
}
\caption{\label{fig:waveforms} (a) Time dependent surface temperature (red curve) induced by the square-modulated (green curve) laser pump (see Appendix \ref{App:TemperatureIncrease}). (b) Schematics of the modulated FM fraction $x\left(t\right)$ (red curve) at pump frequency $f=100$~kHz (green curve) showing the in-phase component  with amplitude $\Delta x^p_f$ (black curve), the maximum, and average values $\Delta x_{\rm{max}}$, $\Delta x_{\rm{av}}$, respectively, and the non-modulated fraction $x_0$.}
\end{figure}

The thermoreflectance contribution is then subtracted from the $V^{\rm{p}}_{\rm{f}}$ and $V^{\rm{q}}_{\rm{f}}$ signals. The remaining modulated reflectance $\Delta R^x_{\rm{f}}$ is assumed to arise from the time modulation of the FM fraction $\Delta x(t)$ depending on the pump power. This corresponds to the fraction of the material that undergoes a phase change due to the periodic temperature pulses induced by the green laser across the spot. Thermal modeling of the film and of its contact with the substrate allows the calculation of the $T$-pulse amplitude at the surface for every power (see Appendix~\ref{App:TemperatureIncrease}). It shows that $\Delta T$ spans a temperature range between a lower temperature $T_{base}$, close to the temperature of the substrate (\textit{i.e.} of the heating stage controller) and slightly departing from it at the highest pump laser powers, and a peak temperature $T_{peak}$ determined by the pump power, $\Delta T = T_{peak} - T_{base}$ as shown in Fig.~\ref{fig:waveforms}(a). The time-evolution of the FM phase over one pump period is schematically depicted in Fig.\thinspace \ref{fig:waveforms}(b). For a given pump power, the time average of the modulated fraction, $\left\langle \Delta x(t) \right\rangle_t=\Delta x_{av}$, and the maximum amplitude of the modulated phase change over a pump period, $\Delta x_{max} = 2 \Delta x_{av}$ represent a measure of the fraction of the material driven back and forth across the phase transition by the periodic temperature pulses of amplitude $\Delta T$. The Fourier component $\Delta x_f$ of $\Delta x(t)$ at frequency $f$, contributes to the detected $f$-harmonic $\Delta R_f$. Using Eq.(\ref{eq:R}) we have:

\begin{eqnarray}
    \Delta R_f&=&\Delta R^{th}_f+ \Delta R^x_f\\
     \Delta R_f^x&=&\frac{dR}{dx}\Delta x_f =(-R_{AF}+R_{FM})\Delta x_f.
\end{eqnarray}
     
For the sake of simplicity we further assume that, after reaching the stationary state, the $\Delta x (t)$ response to a square-wave modulation varies exponentially with the same rise and decay time $\tau$. The relevance of this assumption is discussed in Section~\ref{sec:Nonlinearities}. Hence $\Delta x_{\rm{av}}(T)$ is obtained from the experimental signals as (see Appendix~\ref{App:Deltaxav}):

\begin{equation}
    \Delta x_{\rm{av}}=\frac{\pi \left(1+\left(V_{\rm{f}}^{\rm{q}}/V_{\rm{f}}^{\rm{p}}\right)^2\right)}{2 \sqrt{2}}\frac{R_{AF}}{(-R_{AF}+R_{FM})}\frac{V_{\rm{f}}^{\rm{p}}}{V^{AF}_0}\; ,
    \label{Eq:Deltaxav}
\end{equation}
where $V^{AF}_0$ is the reflectance DC voltage of the AF phase ($T$=50$^{\circ}$C (40$^{\circ}$C) for $P<$13\thinspace mW ($>$13\thinspace mW)).

In spite of the non-linear nature of the signal $x(t)$, $\Delta x_{\rm{av}}$ can be used to estimate the average fraction of phase modulated by the temperature pulses when sweeping a minor thermal hysteresis loop. A further discussion on the role of non-linearity  will be given in Appendices\thinspace \ref{App:Deltaxav} and \ref{sec:Nonlinearities}.

The complex behavior exhibited by $\Delta x_{av}$ as a function of the pump power, and of $T_{base}$ can be understood only by taking into account the thermal hysteresis exhibited by the relative phase fraction. While when  $\Delta T$ lays outside the phase-coexistence temperature interval no phase modulation is expected, when the film temperature comes to cross the phase coexistence region, the phase fraction is expected to follow the temperature variation, and to vary between the value $x_0$, corresponding to the FM fraction at the temperature $T_{base}$, and $x_0 + \Delta x_{max}$. However, hysteresis makes the value $x_0$ at a given time $t_0$ depend on  $T_{base}$ as well as on the temperature history, namely on the values $T(t)$ assumed for $t < t_0$.

In order to get more insight into the effect of the pump laser in the AF-FM mixed phase, we shall hereafter examine in detail the temperature dependence of the reflectance without laser pump within the quasi-static regime (Fig. \ref{fig:Fig1}(c,d)).

\subsection{\label{sec:ReturnPointMemory} Quasi-static and laser modulated thermal hysteresis}

\begin{figure}[ht]
\includegraphics[width=0.98\columnwidth]{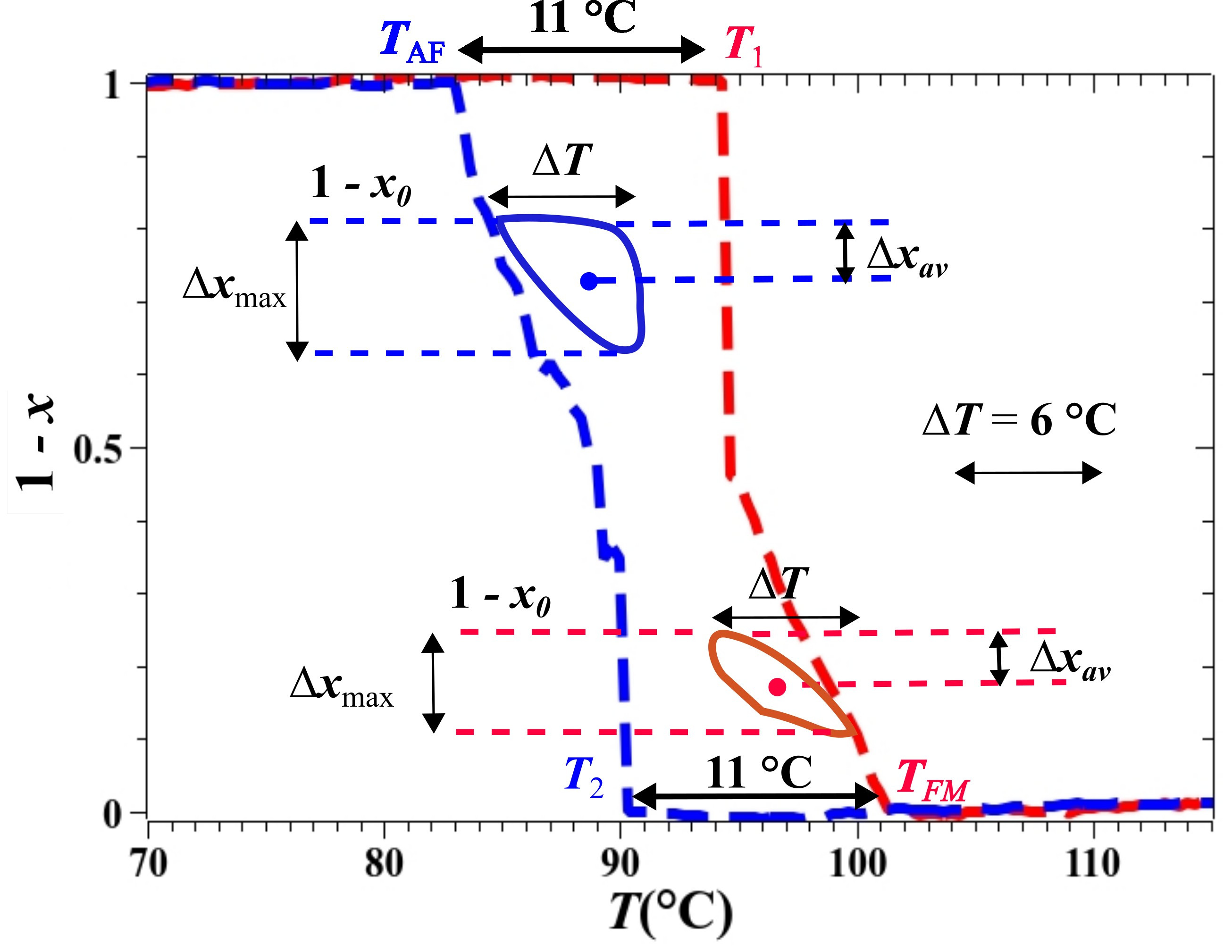}
\caption{\label{fig:Scheme-minor} Dashed curves: AF fraction obtained from the DC reflectance without pump. Schematics of the major thermal hysteresis loop with two first-order minor loops that take place under heating (red curve) and cooling (blue curve) of the heating stage (\textit{i.e.} loops with one vertex lying on the major cycle). The relationship between the quantities $\Delta x_{av}$, $\Delta x_{max}$ measured with the lock-in amplifier under laser pumping when the pump induces periodic $\Delta T = T_{peak} - T_{base}$, and the shape, and the location of the minor loop swept by $\Delta T$ are shown. While the amplitude of the AC signal measured at a given $T_{base}$ is strictly related to the shape of the minor loop, the value of the DC signal depends on the location of the minor loop centre at $(1 - x_0) - \Delta x_{av}$.}
\end{figure}

Since one of their earliest formulation by Madelung \cite{Madelung1905-1, Brokate2012-1, Pierce2007-1}, the general hysteresis properties offer a tool to investigate the way the memory of a physical system is stored, and erased when a driving parameter, temperature in the present case, is changed. The existence of a major hysteresis loop, namely a limit cycle reached by the system when some threshold values of the driving parameter are overcome, and a significant compliance with Return Point Memory (RPM), the property of restoring the microscopic state of the system along the major, and any minor hysteresis cycles \cite{Bertotti1998, Keim2019-1} have been reported on FeRh films \cite{Keavney2018-1, Baldasseroni2012-1}. Measuring thermal hysteresis with the method used here allows the study of the local properties of different spots over the sample. While the temperatures at which the phase coexistence region begins, over heating or cooling, may slightly change depending on the location over the film surface, the shape, amplitude, and width of the hysteresis loops are quite similar everywhere (see Appendix\thinspace \ref{App:Map}). In fact,  general thermal hysteresis properties are uniform across the sample. Regardless of where the reflectance measurements are made, a limit, reproducible major hysteresis cycle can be identified. This implies that two temperatures exist, $T_{AF}$, and $T_{FM}$, as shown in Fig.~\ref{fig:Scheme-minor}, such that when the material is cooled (heated) to a temperature $T \leq T_{AF}$ ($T \geq T_{FM}$) it always gets in the pure AF (FM) state and all its previous memory is erased. In addition, any monotonous heating (cooling) starting at a temperature $T \leq T_{AF}$ ($T \geq T_{FM}$)  brings the system to the major heating (cooling) branches depicted as a dashed red (blue) line in Fig.~\ref{fig:Scheme-minor}. Besides, two temperatures are defined, $T_1$ ($T_2$), where phase coexistence begins when heating (cooling) along a major branch. It is worth noting that, as the reflectance $R$ (\textit{i.e.} the directly measured quantity) scales with the AF phase fraction, in what follows all the hysteresis curves will be plotted as a function of the AF fraction, namely of $1 - x$.

Figure~\ref{fig:Scheme-minor} shows  the schematic quasi-static limit cycle measured over the spot where most of the data discussed here have been acquired (dashed blue and red lines). The four above defined temperatures that uniquely characterize the quasi-static major cycle, $T_{AF}$, $T_1$, $T_{FM}, T_2$, are marked on the figure. Both major branches show, everywhere over the sample,  quite a steep initial part (\textit{i.e.} about half of the overall reflectance change), followed by a smoother section eventually reaching the single phase region. This can be interpreted as due to a phase change starting with the defect-driven nucleation of a large domain of the new phase, followed by the nucleation and coalescence of smaller regions. A similar behavior has been recently observed using X-ray photoemission electron microscopy (X-PEEM) \cite{Keavney2018-1, Arregi2023-1}. 

In addition to the existence of limit cycles, at the scale of the laser spot, when the temperature is quasi-statically changed, RPM has been systematically verified over different positions on sets of minor loops starting from the major heating and cooling branches using the above described protocol (Fig. \ref{fig:Fig1} (d,c)).

When a system exhibiting RPM is driven by a periodic input, a partial order of the microscopic configurations is uniquely defined by the sequence of extrema of the control parameter regardless of its rate of change \cite{Sethna1993-1, Mungan2019-1}. This state of affairs is referred to as rate-independent \cite{Bertotti1998}. In this case RPM describes the fact that, under a periodic input, the ensemble of states visited by the material is completely determined by the absolute maxima and minima previously attained by the control parameter while, when one of the extrema, either the minimum or the maximum, is overcome some memory gets erased, and the system is driven over another set of states defining, in terms of output, a new minor cycle.

Below, we shall cross-check the behavior observed under the modulating effect of the green laser with the quasi-static hysteresis properties presented above. The main purpose is to investigate to what extent quasi-static hysteresis properties measured at the micrometer size without the pump laser, still govern the FeRh behavior when the temperature is periodically changed at the scale of the laser spot with sub-microsecond temperature pulses. 

It is worth noting that the heating conditions under the effect of the laser pulses differ from those driven by the heating stage controller from at least two respects. Firstly, the timescale of the temperature ramps are quite different, from $4$~$^{\circ}$C/min with the heating stage up to $20$~$^{\circ}$C in a fraction of $\upmu$s with the pump laser (see Appendix \ref{App:TemperatureIncrease} for more details). Secondly, the heating stage controller is quasi-statically driving the temperature of the whole film; on the contrary, the green laser pulses are acting just over the observed spot. This makes the measurements of the AC component a picture of the system in a periodically-driven non-equilibrium stationary state \cite{Koyuk2019-1}. 

Assuming that the very same hysteresis properties (\textit{i.e.} rate independence and RPM) govern the phase fraction $x$ under the effect of the temperature pulses forced by the pump laser, the relationship between the static thermal hysteresis cycle and the above defined quantities $x_0$, $\Delta x_{av}$, and $\Delta x_{max}$ is expected to be the one schematically shown in Fig.~\ref{fig:Scheme-minor}.

Any reversal of the temperature change taking place over the major heating (cooling) branch in the interval $T_1 < T < T_{FM}$ ($T_{AF} < T <T_2$) brings the system on a minor hysteresis loop as the one depicted in red (blue) in Fig.~\ref{fig:Scheme-minor}. Both minor loops shown in Fig.~\ref{fig:Scheme-minor} span a $\Delta T = 6$~\textdegree C, like the one that would be swept by setting the pump laser at a power of $\approx~2$~mW. The modulated phase fraction $\Delta x_{max} = 2 \Delta x_{av}$ deduced from the AC signal amplitude are highlighted and represent the vertical span of the minor loop. The DC signal is the value $(1 - x_0) - \Delta x_{av}$ representing the position of the average of $x$ over the whole minor loop.

The actual temperature history of the material, over the spot, is composed of the slowly varying $T_{base}$ jointly with the $\Delta T$ pulses. This means that during a heating ramp starting from $T_{AF}$ where the temperature of the heating stage controller $T_{AF}$=$T_{base}$ with the laser pump on, the absolute maximum temperature recorded in the history is $T_{base} + \Delta T = T_{peak}$. On the other hand, during a cooling ramp starting from $T_{FM}$ the absolute minimum kept in the history is $T_{base}$. Thence, assuming a rate-independent response, during heating the system is expected to be brought to a state, over the heating branch, determined by the extrema of the $T$-history, namely by the absolute minimum $T_{AF}$, the relative one $T_{base}$, and the maximum $T_{peak}$. Similarly, during a cooling ramp starting from $T \geq T_{FM}$, the state of the system is fully defined by $T_{FM}$ (\textit{i.e.} the absolute maximum), the relative maximum $T_{peak}$, and the absolute minimum  $T_{base}$. 

\begin{figure}
\centering
		\includegraphics[width=0.98\columnwidth]{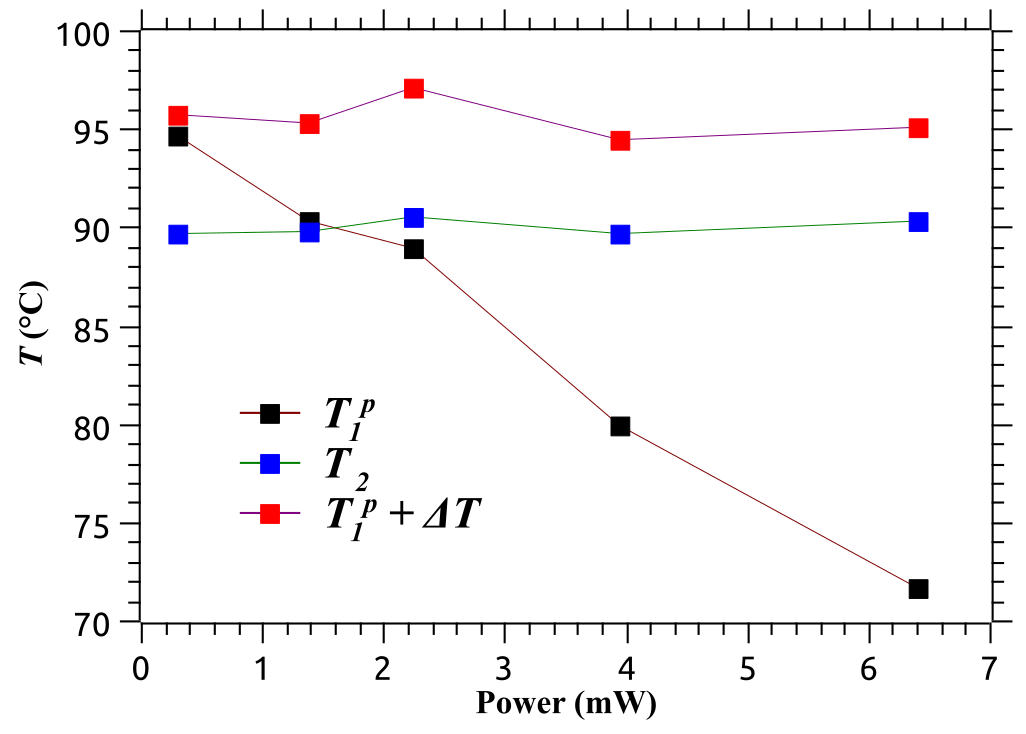}
\caption{\label{fig:T-Power} Pump power dependence of the temperatures at which the DC signal starts decreasing when heating ($T_1^{p}$, black symbols), and increasing when cooling ($T_2^p$, blue symbols). The red points show $T_1^{p} + \Delta T$. }
\end{figure}

A first signature of such a rate-independent, RPM picture can be inferred from the apparent hysteresis reduction of the DC signal under different pumping powers (Fig.~\ref{fig:DC-AC-R}(a)-(e)). Indeed, though the magnitude of the reflectance in AF and FM phases remains unchanged, modulation by the pump laser does not only result in a global shift of the DC-signal hysteresis curve but also in a modification of its shape. Let us name $T_1^p$ ($T_2^p$) the temperature where the DC signal starts decreasing (increasing) over the heating (cooling) ramp, measured when the laser pump is on. While $T_1^{p}$ moves down with increasing pump power as shown in Fig.\thinspace\ref{fig:T-Power}, the temperature $T_2^p$ hardly changes and stays rather close to the pump-free value, namely $T_2 \approx T_2^p$. Thence, the apparent reduction of the thermal hysteresis measured through the DC reflectance under increasing pump power is due to the reduction of $T_1^p$ and to the consequent collapse of the heating curve over the cooling one.

The constant $T_2$ temperature with pump laser hints to nucleation of the AF phase at the layer-substrate interface, which cools down to the heating stage temperature during every OFF half-period of the pump modulation. Note that the reversibility of all the features observed under the pump laser application has been verified by measuring repeatedly the hysteresis properties of the material after removal of the green pump laser.

This behavior is somewhat a natural consequence of the rate-independent, RPM picture. As shown in Fig.~\ref{fig:Scheme-minor} the DC signal represents the central location, $( 1 - x_0) - \Delta x_{av}$, of the minor loop swept by the pulses $\Delta T$. As the DC amplitudes shown in Fig.~\ref{fig:DC-AC-R} are plotted as a function of the temperature of the heating stage (\textit{i.e.} as a function of $T_{base}$) their values start decreasing when $T_{peak} \geq T_1$. Due to the rather abrupt slope of the first part of the major hysteresis cycle, $x_0$ falls towards lower values as soon as $T_{peak} \approx T_1$ bringing the minor loop swept by the $\Delta T$ pulses towards lower locations on the heating branch as made clear by the red minor loop shown in Fig.~\ref{fig:Scheme-minor}. 

Therefore, the DC signal hysteresis reduction under increasing pump power is a by-product of plotting the DC signal as a function of $T_{base}$. Indeed, as long as temperature changes driven by the green laser are led by  the quasi-static hysteresis behavior, phase coexistence is expected to start when $T_{base} + \Delta T = T_1 $ and to be detectable through an increasing modulated reflectance signal (\textit{i.e.} a non-vanishing $\Delta x_{av}$). Indeed, Fig.~\ref{fig:T-Power} shows that $ T_1^p + \Delta T \approx T_1$ is constant as a function of the pumping power. 

\begin{figure}[h!]
\includegraphics[width=0.71\columnwidth]{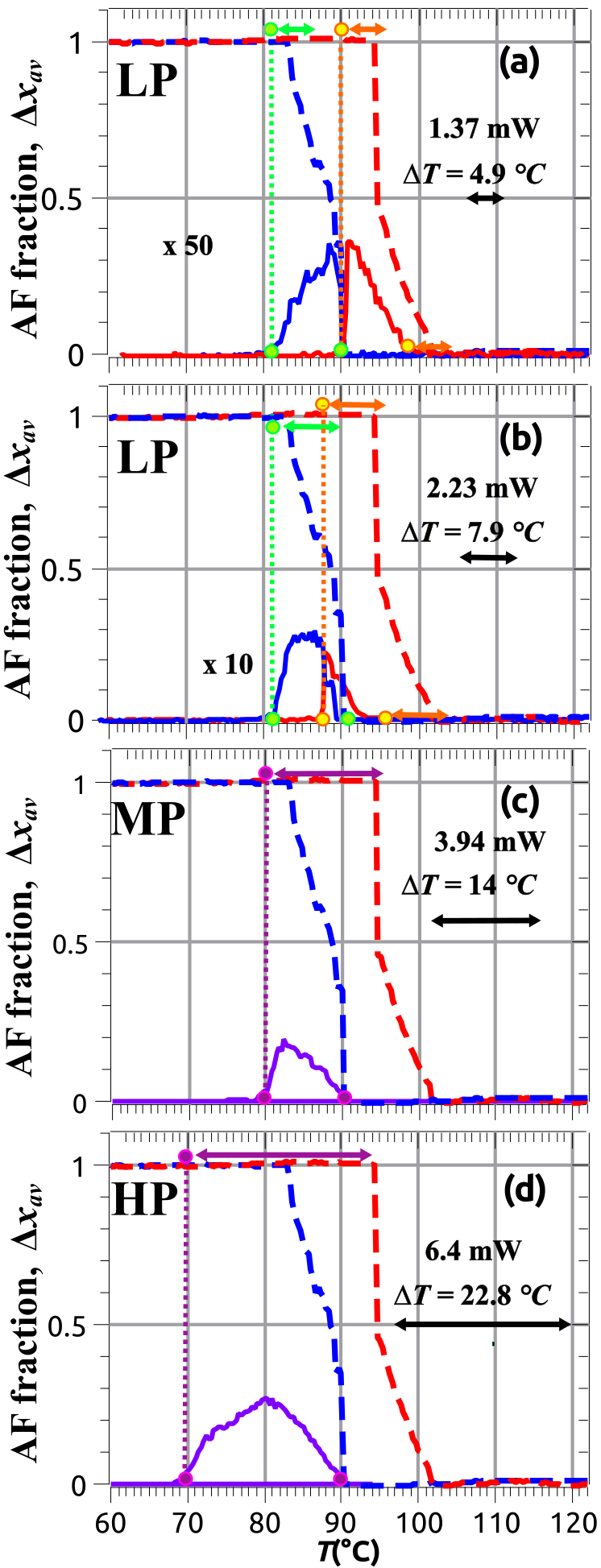}
\caption{\label{fig:XandDeltaX} Dashed curves: AF fraction obtained from the DC reflectance without pump (red for heating and blue for cooling).  Full curves: average modulated fraction $\Delta x_{\rm{av}}$ (red for heating and blue for cooling). The double arrow length is equal to the $\Delta T$ amplitudes. In (a), and (b) (LP) the orange arrows/points show the offset of $\Delta x_{av}$ increasing on heating, $T_{peak} > T_1$, and the point where it vanishes $T_{peak} > T_{FM}$. Green arrows/points show the onset of $\Delta x_{av}$ increasing on cooling, $T_{base} < T_1$, and point where it vanishes $T_{base} < T_{AF}$. Similarly, for the MP, and HP regimes, magenta arrows/points mark the region where $\Delta x_{av} > 0$ and the relative threshold temperatures.}  
\end{figure}

Another feature revealing the leading role played by the quasi-static, pump-free, hysteresis on the modulated amplitude $\Delta x$ can be made clear by discussing the dependence of $\Delta x_{av}(T_{base})$ on $\Delta T$. The $\Delta x_{av}(T_{base})$ amplitude calculated from the measured AC signals are plotted on Fig.~\ref{fig:XandDeltaX} for different values of the pump power (\textit{i.e.} for different $\Delta T$). Three regimes can be defined through the temperature intervals $(T_1 - T_{AF}) \simeq (T_{FM} - T_2) \simeq 11$~\textdegree C, and $(T_{FM} - T_{AM}) \simeq 18$~\textdegree C, characterizing the quasi-static major hysteresis cycle:

\begin{itemize}
    \item Low power (LP):  $\Delta T < T_1 - T_{AF}$;
    \item Medium power (MP):  $T_1 - T_{AF} \leq \Delta T < T_{FM} - T_{AF}$;
    \item High power (HP):  $\Delta T \geq  T_{FM} - T_{AF} $.
\end{itemize}

\begin{figure}[h]
\includegraphics[width=0.75\columnwidth]{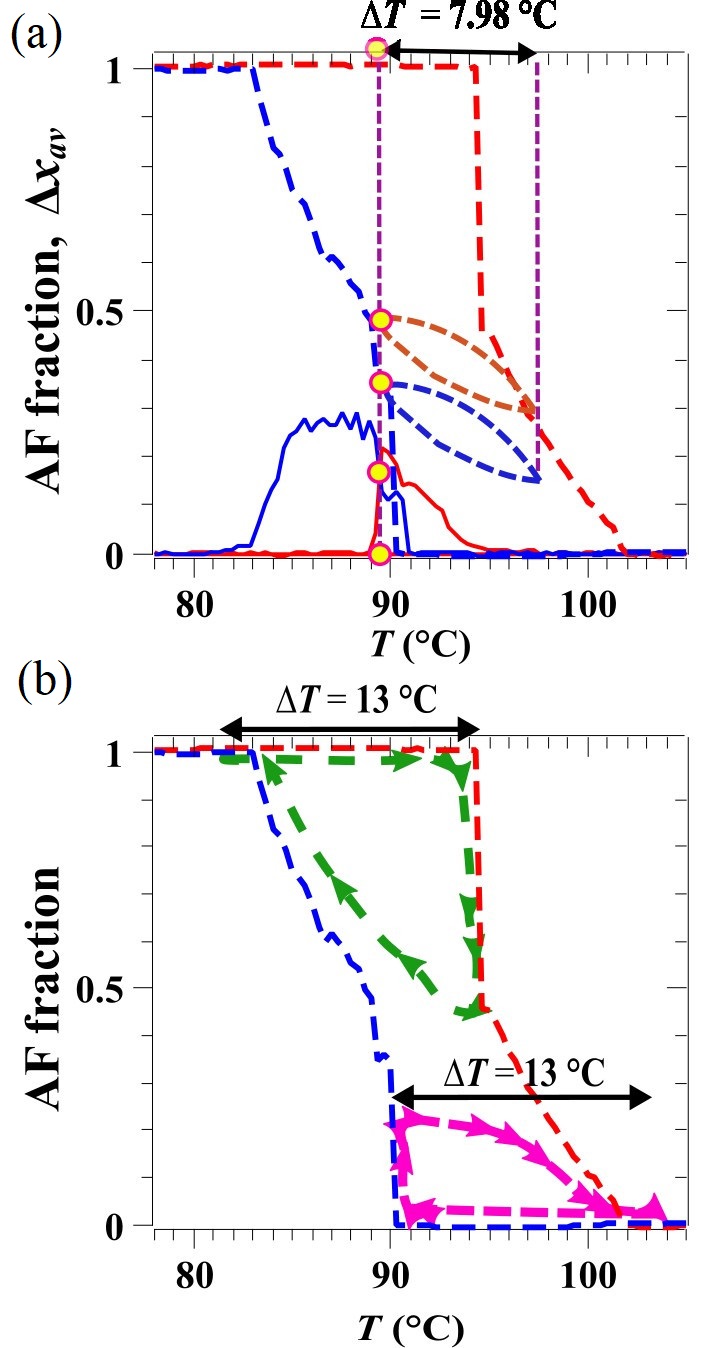}
\caption{\label{fig:return-congruency}Dashed curves: AF fraction obtained from the DC reflectance without pump.(a) Two minor loops swept by the temperature pulses $\Delta T$ under heating, and cooling at the same $T_{base}$ (indicated by the yellow points) in the LP regime superimposed with the $\Delta x_{av}$; (b) the typical return curves where the phase modulation takes place under heating (dashed green curve), and cooling (dashed purple curve) in the MP regime.}  
\end{figure}

Figure~\ref{fig:XandDeltaX} shows $\Delta x_{\rm{av}}(T)$ determined from Eq.\thinspace (\ref{Eq:Deltaxav}) against the major hysteresis cycle exhibited by the AF fraction ($1-x$) measured when the laser pump is switched off. Frames (a,b) show data measured with the pump power within the LP interval, while frames (c) and (d) represent MP and HP ranges, respectively. In the LP interval, where $\Delta T < 11$~\textdegree C,  two well separated $\Delta x_{av}$ signals are apparent, one (red) representing the modulated phase during the heating ramp (\textit{i.e.} minor loops attached to the heating major branch), the other (blue) showing the AC signal measured along the cooling process (\textit{i.e.} minor loops starting from the cooling major branch). The two signals may cross, as apparent in Fig.~\ref{fig:XandDeltaX}(b) due to the superposition of $\Delta T$ on heating, and cooling, but they  remain well apart one from another. This is due to the fact that, even when the two curves coexist within the same temperature interval, phase modulation takes place over different minor loops. The $T_{base}$ where the heating, and cooling $\Delta x_{av}$ curves cross is the point where the two minor loops are congruent,  as depicted in Fig.\ref{fig:return-congruency}(a). 

Instead, within the MP, and HP intervals the $\Delta x_{av}$ measured under heating, and cooling merge into the same curve, as apparent in Fig.~\ref{fig:XandDeltaX}(c, d). In the MP interval, where $\Delta T > (T_1 - T_{AF}) \simeq (T_{FM} - T_2)$, this is due to the fact that phase modulation takes place mostly over first-order return curves like the ones schematized in Fig.~\ref{fig:return-congruency}(b). For instance, during the heating process, when phase modulation starts, at $T_{peak} > T_1$, the minimum of the  temperature oscillating over a pulse lies below $T_{AF}$ (\textit{i.e.} $T_{base} \leq T_{AF}$) so that the temperature pulse is sweeping a return curve similar to the green one schematically depicted in Fig.~\ref{fig:return-congruency}(b).  Similarly, during the cooling process, when $T_{base}$ gets below $T_{AF}$ the temperature pulses  take place over the very same return curve. Analogously, during the heating process when $T_{peak} > T_{FM}$ and $T_{base} < T_2$, under the action of the periodical $\Delta T$ oscillations the material cools from $T_{peak}$ down to $T_{base}$ over the major cooling branch, thereafter it heats up to $T_{peak}$ again following a first-order reversal curve similar to the magenta one shown in Fig.~\ref{fig:return-congruency}(b). The very same area of the cycle is swept during the cooling ramp when $T_{base}$ falls below $T_2$. Therefore, in the MP regime, $\Delta x_{av}$ is mostly the result of phase modulation over the same curves under heating and cooling. 

In the HP interval, when $\Delta T > T_{FM} - T_{AF}$, beside spanning some sets of first-order reversal curves like in the MP region,  the pulses are sweeping the full major hysteresis cycle. Either way $\Delta x_{av}$ results from phase modulation taking place over identical thermal paths under heating and cooling.

It is worth noting that the temperature interval where $\Delta x_{av}$ is different from zero is completely determined by the extrema of the temperature history, namely by the minimum $T_{base}$, and the maximum $T_{peak}$ and by the temperatures $T_{AF}$, $T_1$, $T_{FM}$, $T_2$ defining the phase coexistence intervals on the quasi-static limit hysteresis cycle. This shows that, at least under stationary cyclic driving of the temperature, the rate-independent, quasi-static hysteresis features are leading the material's response.

This state of affairs is illustrated in Fig.~\ref{fig:XandDeltaX}. When measuring the modulated reflectance, at fixed pump power (\textit{i.e.} $\Delta T = $ const.), with $\Delta T \leq T_{FM} - T_{2}$, along a monotonic increasing of $T_{base}$ starting from the single phase region, $\Delta x_{av} = 0$ when $T_{peak} < T_1$, or when $T_{peak} > T_{FM}$ and $T_{base} > T_2$. These conditions are always verified, as highlighted in Fig.~\ref{fig:XandDeltaX}(a,b) with the help of the orange arrows/points. On the other hand, along the cooling ramp $\Delta x_{av} = 0$ whenever $T_{base} > T_2$, or $T_{base} < T_{AF}$ and $T_{peak} < T_1$ as apparent from the green arrows in  Fig.~\ref{fig:XandDeltaX}(a,b). Furthermore, the magenta points, and arrows in Fig.~\ref{fig:XandDeltaX}(c,d) show that in the MP, and HP regime $\Delta x_{av}$ vanishes whenever $T_{peak} < T_1$, or $T_{base} > T_2$.  This confirms that under the effect of pump laser temperature pulses, thermal hysteresis over the spot is led by the very same threshold temperatures that have been identified when heating and cooling quasi-statically the whole sample using the heating stage.

As a further check of the role of RPM, and in general of memory persistence/erasure mechanisms underlying the behavior observed under laser pumping, in the following section a rate-independent hysteresis model showing RPM is used to qualitatively reproduce the measured results.

\section{Modeling}\label{sec:model}

\begin{figure}[ht]
\includegraphics[width=0.9\columnwidth]{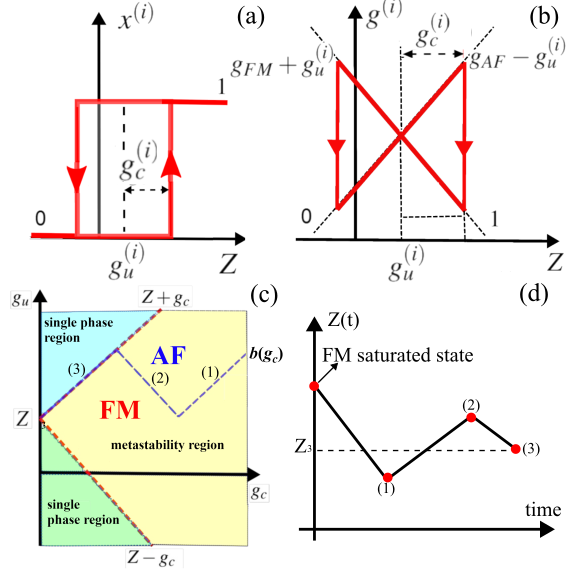}
\caption{\textbf{(a)}: elementary hysteresis loop associated with the $i$ bistable unit; (b) specific Gibbs free-energy of the elementary bistable unit $i$. In both frames, (a), and (b) the switching thresholds of $Z$, from AF to FM at $Z = g_u^{(i)}+g_c^{(i)}$, and from FM to AF at $Z = g_u^{(i)}-g_c^{(i)}$ are shown. (c) the $(g_c, g_u)$ plane. At a given value of $Z$ all the units laying above the $Z + g_c$ (positive slope red dashed line), and below the $Z - g_c$ (negative slope red dashed line) lines are in the single AF, and FM phase respectively. On the contrary, the state of the units within the metastability region (yellow area) depends on the $Z$ history  through the shape of the $b(g_c)$ line. All the bistable units lying above (below) the line $b(g_c)$ (blue dashed line) are in the AF (FM) state. The shape of the line represents the memory of the $Z$ history. The latter, in the case considered here, of rate-independent hysteresis, is uniquely defined by the sequence of extrema reached by $Z$ all along its history; in (d), the extrema defining the $b(g_c)$ shown in (c) are shown.}
\label{fig:preisach-schem}
\end{figure}

In what follows, a hysteresis model complying with return point memory will be used to reproduce qualitatively the main features observed on FeRh films under the periodical temperature driving induced by the pump laser. The simplest mathematical model exhibiting RPM is the one proposed by Ferenc Preisach in $1935$ \cite{Preisach1935-1} based on the assumption of the existence of an ensemble of non-interacting bistable units each one exhibiting different values of the free-energy minima, and of the energy barrier separating them. The model has been successfully applied by Néel to describe the magnetization process in the Rayleigh region \cite{Neel1942-1}. Actually, in spite of the extremely simple assumptions underlying it, its ability to describe some very general features of memory formation and erasure in matter has raised an interest never extinguished for many decades \cite{Biorci1958-1, Wohlfarth1964-1, Porteseil1977-1, Mayergoyz1986-1, Sethna1993-1, Keim2019-1, Mungan2019-1}. 

\begin{figure}[ht]
\includegraphics[width=1\columnwidth]{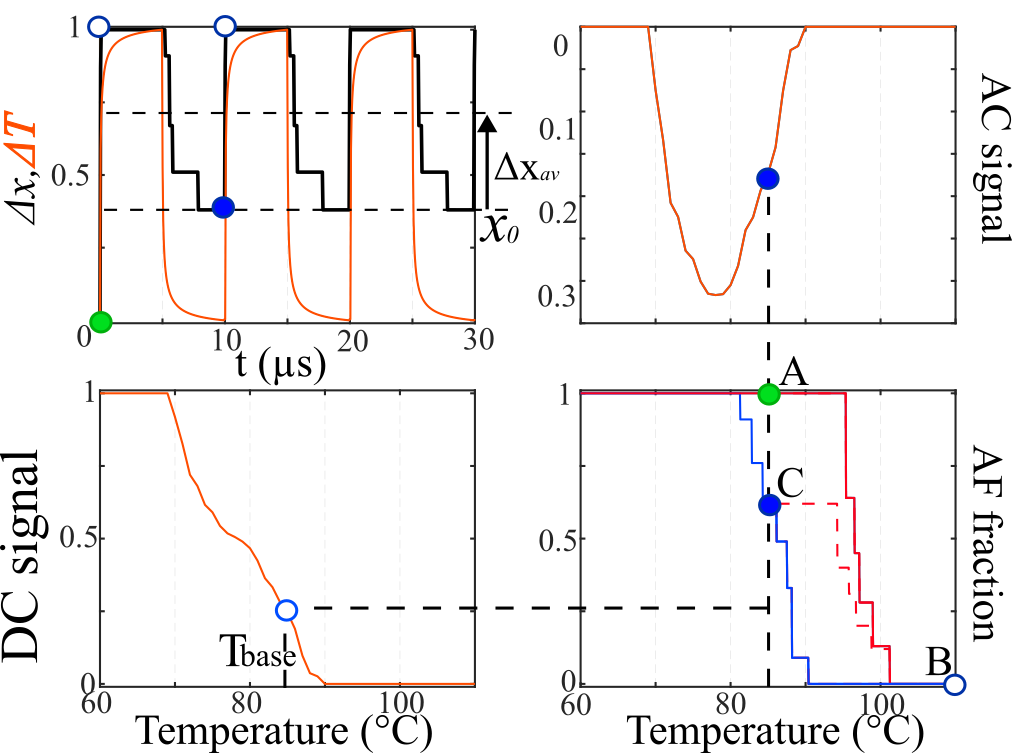}
\caption{(a) Temperature input used in the model (orange line) together with the model output $x(t, T)$ (black line) representing the phase modulation when $T_{base} = 85$\textdegree C; (b) AC signal $\Delta x^p_f$ obtained from the harmonic analysis of the model output $x(t, T)$; (c) DC signal obtained as the average over the model output $\langle x(t, T) \rangle$; (d) the actual output of the model, in terms of $1 - x(T)$ (\textit{i.e.} AF fraction) corresponding to a sequence of $\Delta T = 25$\textdegree C pulses (\textit{i.e.} HP regime) starting from the point A within the pure AF phase region. The first temperature pulse irreversibly brings the material from A (full green point in (a), and (d)) to B (hollow blue point in (a), and (d)), in the pure FM phase region. Afterwards, when $T$ is reduced back to the initial temperature $T_{base}$, it follows the major cooling branch to C (full blue point in (a), and (d)). All the following heating, and cooling curves take place on the red dashed curve C$\rightarrow$B, and on the major cooling branch B$\rightarrow$C respectively giving as an output the AC signal shown in (b) (full blue point), and the DC signal shown in (c) (full orange point).}
\label{fig:model-results}
\end{figure}

Here, a Preisach model recast to describe the thermodynamics of first-order phase transitions with hysteresis \cite{Basso2007, Bertotti2006} is used. Starting from the so-called \textit{local bistability hypothesis}, the existence of a coarse reticulation of the system where each cell contains a single phase volume of the material \cite{Planes1992} is postulated. In this way the cells behave as independent elementary bistable units to be found in one of the two phases, either the AF or the FM in the case considered here. Thence, the local FM phase fraction $x^{(i)}$ at the level of a single cell $i$ is $x^{(i)} = 1$ when the unit is in the FM phase, and $x^{(i)}= 0$ when it is in the AF phase (Fig.~\ref{fig:preisach-schem}(a)). The overall FM fraction is given by the sum, or the integral over all the elementary units at a given moment. Given the Gibbs specific free energies of the two pure phases, $g_{AF}(T, Y_1, ... Y_n)$, and $g_{FM}(T, Y_1, ... Y_n)$, where $Y_i$ are the relevant intensive variables (\textit{e.g.} magnetic field $H$, hydrostatic pressure $p$, stress $\sigma$, etc.), the driving force acting on the phase transformation is assumed to be $Z = (g_{FM} - g_{AF})/2$. In this way, hysteresis and metastability are described by associating with each cell $i$ two parameters $g_c^{(i)}$, and $g_u^{(i)}$ defining, through the following switching rules, the way the unit $i$  behaves given the value of the control parameter $Z$, and its history,

\begin{equation}\label{eq:switching}
\begin{split}
    Z \geq g_u^{(i)}+g_c^{(i)} \; \rightarrow \; x^{(i)} = 1 \\
    Z \leq g_u^{(i)}-g_c^{(i)} \; \rightarrow \; x^{(i)} = 0.
\end{split}
\end{equation}

As apparent from Fig.~\ref{fig:preisach-schem}(a,b), $g_c^{(i)}$ can be interpreted as the width of the elementary hysteresis cycle associated with the $i$ unit, and as the energy barrier to be overcome for switching from one metastable state to the other. On the other hand $g_u^{(i)}$ defines the asymmetry of the elementary bistable with respect to $Z$, namely the free-energy difference between the two energy minima of the bistable unit. Given the value of $Z$, these switching rules allow defining, over the $(g_c, g_u)$ plane, a metastability region,  $g_u^{(i)}-g_c^{(i)} < Z < g_u^{(i)}+g_c^{(i)}$, where the state of the elementary cell, in one or another of the phases, depends on the previous $Z$ history. In addition, the same rules define a border $b(g_c)$ over the  plane, separating the units in the FM state from the units in the AF one. The shape of the $b(g_c)$ line contains the full information on the memory of the system. A picture of  the $(g_c, g_u)$ plane for a given value of $Z$ is shown in Fig.~\ref{fig:preisach-schem}(c) where the shape of the $b(g_c)$ line represents the memory of the previous $Z$ history. The overall FM phase fraction $x$ can be calculated as the superposition of the state of the bistable units as follows:

\begin{equation}
    x = \int_0^\infty d g_c \int_{- \infty}^{b(g_c)} p(g_c, g_u)\; d g_u\; ,
\end{equation}

\begin{figure}[ht]
\includegraphics[width=0.9\columnwidth]{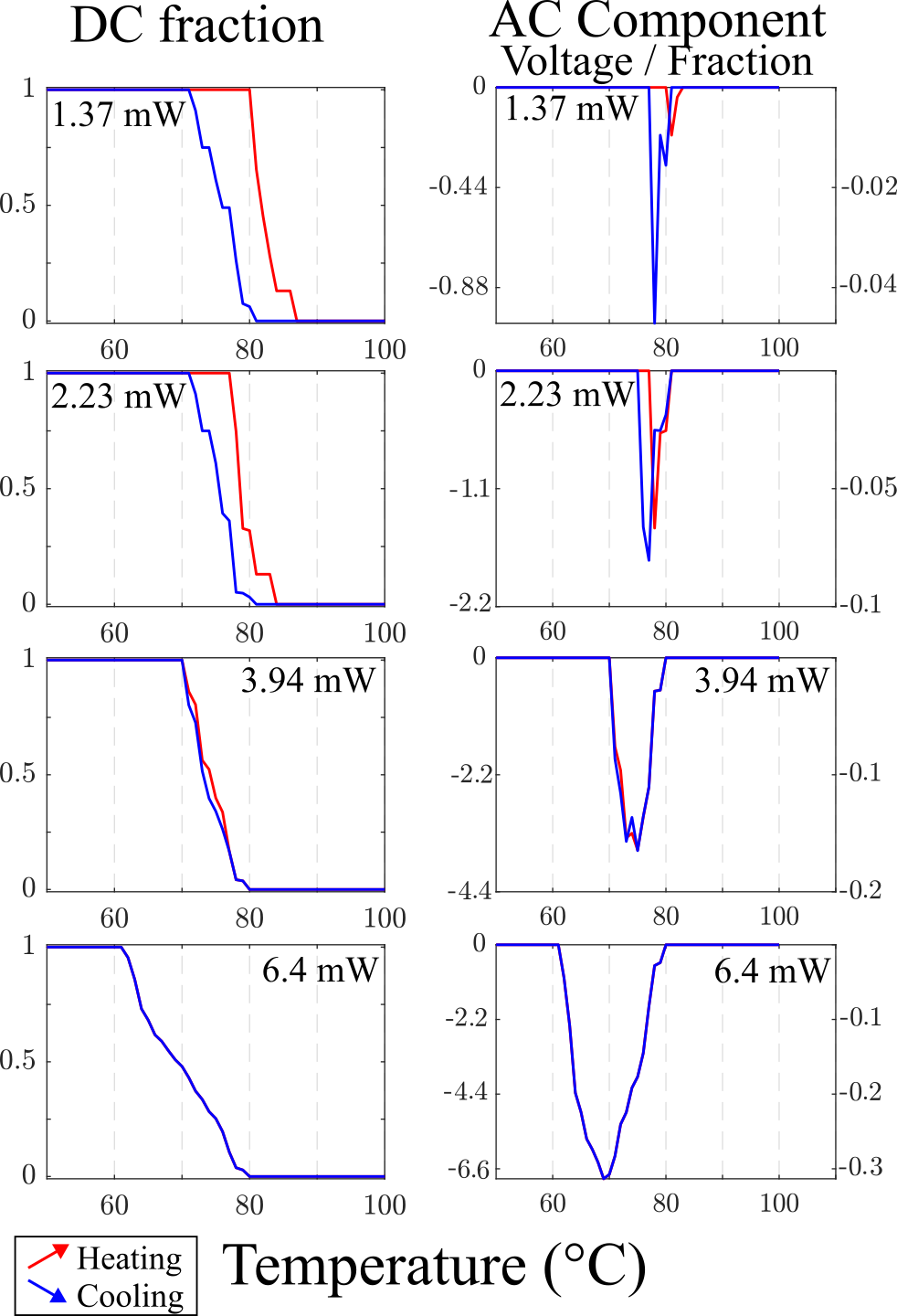}
\caption{ DC, and AC signal obtained through cycling temperature using a rate-independent Preisach model as defined in Appendix \ref{App:Hysteron-Model}. The curves match rather well the ones obtained using the lock-in amplifier shown in Fig.\ref{fig:DC-AC-R} after subtraction of the thermoreflectance signal $\Delta R^{th}.$ The AC signal (right column) is plotted as a funtion of the relative phase fraction (left axis in the diagrams). The right axis shows the relative value rescaled by $\Delta V = 22$~mV in order to make the comparison with the measured AC signal shown in Fig.\ref{fig:DC-AC-R} easier.}
\label{fig:toy-model-AC-DC}
\end{figure}

where the function $p(g_c, g_u)$ is the bistable distribution, a property describing the way defects, nucleation centres, and in general quenched disorder determine the hysteresis properties of a given material. Numerical methods to identify $p(g_c, g_u)$ from a set of minor loops, or using first-order reversal curves have been described in \cite{Mayergoyz2003}. The interpretation of numerically reconstructed bistable units distributions has often been discussed as a fingerprint of the material properties \cite{Cornejo1997, LoBue1998, Pike2003-1, Zhou2010-1, Komlev2021-1}. Using such methods would be beside the point here, since our primary objective is to show that, despite a detailed reproduction of hysteresis curves, an extremely coarse hysteresis model based on the identification of a finite number of elementary units is able to grasp the main qualitative features of phase modulation thanks to its rate-independent nature, and to the RPM property. To this end, a $10^2$ units model has been formulated using the identification protocol described in Appendix \ref{App:Hysteron-Model}.

The $\Delta T$ pulses at a given $T_{base}$ have been used in the model to calculate the resulting phase modulation. An example of the $T$ input and of the calculated $x(T)$ output are shown in Fig.~\ref{fig:model-results}(a) (orange and black full lines, respectively) where $\Delta T = 25$\textdegree C (\textit{i.e.} HP regime), and $T_{base} = 85$\textdegree C. Fig.~\ref{fig:model-results}(d) shows the return curve swept  by the periodic pulses. The average over the modulated phase shown in Fig.~\ref{fig:model-results}(a) allows calculating the DC signal at $T_{base}$, shown as an orange point in Fig.~\ref{fig:model-results}(c). The harmonic analysis of the modulated signal is used to extract the value of the AC signal at the same $T_{base}$ depicted as a blue point in Fig.~\ref{fig:model-results}(b). 

The same protocol is used to calculate the model response to a periodic temperature forcing at different $\Delta T$ corresponding to the three power regimes defined above. A unique $\Delta T$ has been used since the thermal gradient within the film depth calculated from the model has been found to be negligible. In reality, it never overcomes $10 \%$ of $\Delta T$ upon heating and is almost zero during the cooling phase (\textit{i.e.} after less than $1~\upmu$s). The calculated DC component, shown in Fig.~\ref{fig:toy-model-AC-DC}, is in very good agreement with the data presented in Fig.~\ref{fig:DC-AC-R}, and shows all the features presented above with an almost complete curve merging after $3.94$~mW. The AC signals plotted in Fig.~\ref{fig:toy-model-AC-DC} reproduce qualitatively the shape of the measured curves, and their evolution as a function of the power (\textit{i.e.} shape and ratio between the AC component at different powers). The right axes of the AC component show the normalized values calculated from the harmonic analysis of the model. The left axis is rescaled to the measured voltage (\textit{i.e.} $\Delta V = 22$~mV). The slight overestimation of the absolute value of the AC signal by the model is to be ascribed to the simplified identification approach used to define the bistable distribution.

\section{\label{sec:Conclusion}Conclusion\protect}

A set of measurements carried out with a modulated thermoreflectance setup have been presented. Beside the thermoreflectance $\Delta R^{th}$, studied elsewhere \cite{Castellano2024} for extracting the thermal properties of the sample, an additional signal, $\Delta R^x$, present within the FeRh phase coexistence temperature interval, and depending on the power of the pump laser, is singled-out and studied. 
Using a thermal model the time-dependent temperature induced by the pump laser over the material surface is worked out, and the modulated reflectance signal is shown to be associated with the time-dependent FM phase fraction $x(t)$ driven by the periodic temperature pulses induced by the pump laser. In this way, the modulated phase fraction is used to gain insight into the thermal hysteresis associated with the FeRh AF-FM phase transition under periodic temperature change. 

This allows to study FeRh thermal hysteresis properties under temperature rate of changes that would be hardly achievable using a standard temperature controller. Moreover, while phase domains, and thermal hysteresis have been studied at the micrometer scale by keeping the material at thermal equilibrium with a heating stage \cite{Arregi2023-1, Keavney2018-1, Baldasseroni2012-1}, here measurements are performed over a spot that is periodically driven away from thermal equilibrium with the substrate, as well as with the rest of the FeRh film. 

The quasi-static measurements presented above confirm most of the features reported in the literature. The existence of a major limit cycle, and compliance with RPM at the scale of minor cycle are verified. Besides, while the major cycle shape is symmetric over heating, and cooling, the minor hysteresis loops starting from the heating, and cooling branches are rather different as apparent in Fig.\ref{fig:Fig1}(d). This feature has been repeatedly reported \cite{Maat2005-1, deVries2014-1, Baldasseroni2015-1, Arregi2023-1}, and is generally interpreted as a signature of different mechanisms underlying the nucleation, and kinetics of the AF, and FM phases respectively. These equilibrium measurements are used as a reference to discuss the out-of-equilibrium measurements of the phase fraction modulated by laser pumping.

The main finding here is to show that the very same rate-independent RPM features are still leading the behavior of the material when temperature is locally changed, with average rates up to tens of degrees per microsecond. This means that, in spite of the different kinetics involved in the cooling/heating parts of the minor loop spanned by the pump induced temperature pulses, on the average, the hierarchy of states defining the minor loop remain unchanged over a temperature rate of change spanning several order of magnitudes, namely from $\approx 10^{-1}$~$\textdegree$C/s up to $\approx 10^7$~$\textdegree$C/s. This finding does not exclude the presence of rate-dependent or transient phenomena that may become apparent through a time-resolved measurement. Nonetheless, it shows an outstanding stability, on average, of the phase-coexistence pattern as defined by the quasi-static, rate-independent properties of the material. The relevance of the result is manyfold. On the one hand, the stability of the AF/FM domain pattern is a key feature for recording, and patterning applications \cite{Thiele2003-1, Marti2014-1, Mei2020a}. On the other, the result reveals a stability, and robustness of the energy landscape tailored by quenched disorder that  confirms the leading role played by the defects, either compositional \cite{Staunton2014-1}, or structural hole-like ones \cite{Arregi2023-1} that guide the phase coexistence patterns over a scale of tens of microns. 

The result also offers a fresh insight into how memory is stored, and erased in disordered materials under cyclic driving conditions. Many recent studies are devoted to this issue focusing on mechanical memory in shared solids \cite{Mungan2019-1, Paulsen2024-1}, as well as on the general features of memory formation in solid state systems \cite{Keim2019-1}. Particularly interesting is the way periodically-driven many-body systems organize into a state where the system visits the same sequence of microscopic configurations over every driving cycle \cite{Reichhardt2023-1}. Indeed, when submitted to a cyclic driving, a system showing hysteresis may spend a certain amount of cycles within a transient training till reaching a stable state uniquely defined by the extrema of the periodic driving. A similar behavior has been studied in magnetic systems submitted to small amplitude oscillating magnetic fields since the works of Néel on magnetic viscosity in the Rayleigh domain \cite{Neel1950-1}. The measurements presented here show that FeRh films submitted to cyclic temperature driving, organize into a stable state showing a periodic response on a timescale short enough to conceal any training, or relaxation. Namely, the system reaches the periodic response regime within the first few driving periods, a timescale long compared to the subpicosecond transient reflectance response that has been recently reported \cite{Harton2024}. The time average of this response, measured with the lock-in amplifier, is stable against the thermal and mechanical fluctuations associated with the coupling with the surrounding medium, a robustness of the cyclic response that is not always observed in disordered media \cite{Paulsen2024-1, Majumdar2023-1}.

This stability over time average deserves further studies. A time-resolved analysis of the phase modulated signal under pumping, beyond the scope of the present work, will allow to understand the role of fluctuations on time averages. Indeed, many random features, that can involve the signal amplitude, as well as its delay with respect to the pump laser induced temperature pulses, are expected to play a role in the time averages. The very same variation of the $\Delta x_{av}$ amplitude can originate from an actual reduction of the amplitude of the signal, or from the stochastic nature of the transition induced by the laser pulses. 

Another feature that will deserve further investigations is the response of the phase fraction to other periodic stimuli, as the periodic strain induced by surface acoustic waves (SAW) \cite{Wu2024-1}.

\begin{acknowledgments}
This work has been partly supported by the French Agence Nationale de la Recherche (ANR ACAF 20-CE30-0027), by the LabEx LaSIPS (ANR-10-LABX-0032-LaSIPS) FLIGHTPATH under 
the program "Investissements d’avenir" (ANR-11-IDEX-0003) and the Ecole Normale Supérieure Rennes. We acknowledge M. Bernard, M. Vabre, and S. Vincent (Institut des Nanosciences de Paris) for technical assistance. Access to the CEITEC Nano Research Infrastructure was supported by the Ministry of Education, Youth and Sports (MEYS) of the Czech Republic under the project CzechNanoLab (LM2023051).
\end{acknowledgments}

\appendix

\section{Pump-induced temperature increase and thermoreflectance signal}\label{App:TemperatureIncrease}

In order to calculate the thermoreflectance contribution to the modulated reflectance, we first need to obtain the $f$-harmonic $\Delta T_f$ of the temperature increase. This is done by solving the heat diffusion equation in cylindrical coordinates in a layer-on-infinite-substrate geometry using the Hankel transform and looking for solutions $\Delta T_f (r,z) \exp(j \omega t)$, where $z$ is the normal-to-the-layer axis, $r$ the in-plane radial coordinate, $\omega$ the angular frequency $\omega=2 \pi f$, and $j=\sqrt{-1}$. At the surface ($z=0$) the temperature increase and the thermoreflectance signal detected by the focused laser probe read:

\begin{widetext}
\begin{eqnarray}
\Delta T_{\rm{f}}\left(r=0,z=0\right)&=&I\left(d\right)\\ \label{eq:DeltaTf}
\rm{with} \quad I\left(d\right)&=&\frac{Q_{f}}{2\pi}\int_0^{\infty} g(u) \;u\;\exp{\left(-\frac{d^2u^2}{32}\right)}\;du\;, \label{eq:I}\\
g(u)&=&\frac{\beta^2}{\beta^2-\sigma^2}\frac{\left(1-\frac{\kappa_s\sigma_s}{\kappa\beta}\right)\left(\cosh({\sigma h})-\exp({-\beta h})\right)+\left(\frac{\kappa_s\sigma_s}{\kappa\sigma}-\frac{\sigma}{\beta}\right)\sinh({\sigma h})}{\kappa_s\sigma_s\cosh({\sigma h})+\kappa\sigma\sinh({\sigma h})}\; , \label{eq:g}
\end{eqnarray}
\end{widetext}
where $\sigma$ stands for $\sigma(u,\omega)=\sqrt{u^2+j\frac{\omega }{D}}$ and is akin to a complex thermal wave-vector. Index $s$ refers to the  MgO substrate. $\kappa$ is the thermal conductivity, $D=\kappa/(\rho C)$ is the heat diffusivity with $\rho$ and $C$ the mass density and specific heat \cite{Richardson1973}, respectively. The temperature dependencies of the thermal conductivity and diffusivity of MgO are taken from \cite{Hofmeister2014}. The factor $Q_f=\frac{4}{\pi}P_{inc}(1-R_p)$ is the $f$-component of the incoming modulated heat power with $R_p$ the reflectivity at the wavelength of the pump laser and $P_{inc}$ the average laser power impinging on the sample. $\beta$ is the absorption coefficient of the pump beam in FeRh, $\beta_{AF}(\beta_{FM})$=1.1 (0.85)\thinspace 10$^8$\thinspace m$^{-1}$ (using an infinite absorption coefficient only results in a 1.5\% difference in the determination of the temperature increase). The Gaussian function is the Hankel transform of the Gaussian laser spot profile, where $d=d_{\rm{pump}}$ is the diameter at which the intensity of the Gaussian pump spot has fallen by $e^2$. The thermoreflectance signal is calculated as
\begin{eqnarray}
\frac{V_{\rm{f}}^{th}\left(T\right)}{V_0^{AF}}=\frac{1}{\sqrt{2}R_{AF}}& \left(\frac{dR_{AF}}{dT}\left(1-x\left(T\right)\right)I_{AF}\left(d_{\rm{eff}}\right)\right. \nonumber\\
&\left.+\frac{dR_{FM}}{dT}x\left(T\right)I_{FM}\left(d_{\rm{eff}}\right)\right)\;,
\end{eqnarray}
where $d_{\rm{eff}}=\sqrt{d_{\rm{pump}}^2+d_{\rm{probe}}^2}$. $R_{AF/FM}$ is the probe laser reflectance in the AF/FM phase ($R_{AF}\approx$0.66, $R_{FM}$=0.965 $R_{AF}$).  $I_{AF/FM}$ is the $I$-function defined from Eqs.\thinspace (\ref{eq:I}, \ref{eq:g}) using the AF/FM parameters for $\kappa$, $D$, $\beta$ . The $\sqrt{2}$ factor accounts for the lock-in detection of the rms value of the
signal. Note that $V_{\rm{f}}^{th}$ is a complex quantity from which the in-phase $V_{\rm{f}}^{\rm{p}}$ and out-of-phase $V_{\rm{f}}^{\rm{q}}$ thermoreflectance signals can be calculated. The parameters $\frac{dR_{AF/FM}}{dT}$ are adjusted to fit the  $V_{\rm{f}}^{\rm{q}}\left(T\right)$ voltage, which is virtually insensitive to the other parameters.  $\kappa_{AF/FM}$ are slightly adjusted around their average values determined in \cite{Castellano2024} ($\kappa_{AF}\approx$25-30\thinspace W\thinspace m$^{-1}$\thinspace K$^{-1}$, $\kappa_{FM}\approx$10\thinspace W\thinspace m$^{-1}$\thinspace K$^{-1}$)  to fit the experimental $V_{\rm{f}}^{\rm{p}}$. The specific heat for determining the diffusivity of FeRh is taken from \cite{Richardson1973} but only the diffusivity of the substrate actually plays a role in the temperature increase. Let us note that the latent heat of the phase transition has been estimated to be three orders of magnitude smaller than the thermal energy delivered by the laser over the cycle, considering an entropy change value of $12$~JK$^{-1}$kg$^{-1}$\cite{Lyubina_2017}.

The stationary temperature increase $\Delta T_{av}$ can also be calculated from Eq.~\ref{eq:DeltaTf} setting $\omega=0$ in $\sigma(u,\omega)$ and taking $Q_0=P_{inc}(1-R_p)$. The maximum temperature increase $\Delta T$ is exactly equal to 2$\Delta T_{av}$ when the modulation period is much longer than the risetime of the temperature. If not, it can be calculated from the time dependence of the temperature 

\begin{widetext}
\begin{eqnarray}
T(t)=T_{base}+\Re\left(-j\frac{ Q_f}{2 \pi} \sum _{p=0}^{\infty } \frac{1}{2p+1}\exp\left(j   \omega_p t\right) \int_0^{\infty } 
   \exp\left(-\frac{ d_{\rm{pump}}^2 u^2}{32}\right)  g(u,p)\; u \, du\right) \label{eq:TimeDependence}
\end{eqnarray}
\end{widetext}

where $g(u,p)$ is the same as in Eq.\thinspace \ref{eq:g} with $\sigma(u,\omega_p)=\sqrt{u^2+j\frac{\omega_p }{D}}$ and $\omega_p=\left(2p+1\right) \omega$. The time dependence $T(t)$ is shown in Fig.\thinspace \ref{fig:waveforms} for $T_{base}$=60\thinspace $^\circ$C and $\kappa$=29\thinspace W\thinspace m$^{-1}$\thinspace K$^{-1}$. The temperature increase $\Delta T$ is then equal to $\left(T\left(\frac{\pi}{\omega}\right)-T(0)\right)$.

\section{Determination of the average modulated FM fraction $\Delta x_{av}$\label{App:Deltaxav}}

We assume an exponential response of the modulated reflectance signal $V(t)$ to the modulation of the pump with a rise time $\tau_r$ and a decay time $\tau_d$. If $\tau_r$ and  $\tau_d$ are not much smaller than the half-period  $T_0 /2$ of the modulation, $V(t)$ does not reach its $t\rightarrow \infty$ value after a half period. Its rise and decay time dependencies $V_{r}\left(t\right)$ and $V_{d}\left(t\right)$, respectively, are then:

\begin{equation}
V_{r}\left(t\right)=V_{m} \frac{\left(1-e^{-\frac{T_0}{2 \tau _d}}\right) \left(1-e^{-\frac{t}{\tau _r}}\right)+e^{-\frac{T_0}{2 \tau _d}} \left(1-e^{-\frac{T_0}{2 \tau
   _r}}\right)}{1-e^{-\frac{T_0}{2}  \left(\frac{1}{\tau _d}+\frac{1}{\tau _r}\right)}}
\end{equation}

\begin{equation}
V_{d}\left(t\right)= V_{m}
   \frac{\left(1-e^{-\frac{T_0}{2 \tau _r}}\right) e^{-\frac{t-\frac{T_0}{2}}{\tau _d}}}{1-e^{-\frac{T_0}{2}  \left(\frac{1}{\tau _d}+\frac{1}{\tau _r}\right)}}\; ,
\end{equation}
where $V_{m}$ is the maximum amplitude in the limit $\tau_r \rightarrow 0$ and $\tau_d \rightarrow 0$.

The average value is:

\begin{equation}
V_{av}= \frac{V_{m}}{2} \left(1+\frac{2 \left(\tau _r-\tau _d\right) \left(1-e^{\frac{T_0}{2 \tau _d}}\right) \left(1-e^{\frac{T_0}{2 \tau _r}}\right)}{T_0 \left(1-e^{\frac{T_0}{2}
    \left(\frac{1}{\tau _d}+\frac{1}{\tau _r}\right)}\right)}\right)\; .
\end{equation}

The in-phase and in-quadrature components at frequency $f=1/T_0$ are calculated as:

\begin{widetext}
 \begin{eqnarray}
V^{\rm{p}}_{\rm{f}}&=&V_{m}\, \frac{2 }{T_0}\left(\int_0^{\frac{T_0}{2}} V_r (t) \sin
   \left(\frac{2 \pi  t}{T_0}\right) \, dt+\int_{\frac{T_0}{2}}^{T_0} V_d(t) \sin \left(\frac{2 \pi  t}{T_0}\right) \, dt\right)\\
V^{\rm{q}}_{\rm{f}}&=&S_{m}\, \frac{2 }{T_0}\left(\int_0^{\frac{T_0}{2}} V_r(t) \cos
   \left(\frac{2 \pi  t}{T_0}\right) \, dt+\int_{\frac{T_0}{2}}^{T_0} V_d(t) \cos \left(\frac{2 \pi  t}{T_0}\right) \, dt\right)\;.
 \end{eqnarray}
\end{widetext}

 Replacing $T_0$ by $T_0/2$ the components at $2f$ are similarly calculated.

The phase at $f$ is obtained from:
\begin{widetext}
\begin{equation}
   \tan\phi\left(\alpha _r,\alpha _d\right)= -\frac{\pi  \left(\left(\alpha _d^2+\pi ^2\right) \alpha _r \sinh \left(\frac{\alpha _d}{2}\right) \cosh \left(\frac{\alpha _r}{2}\right)+\alpha _d \left(\alpha
   _r^2+\pi ^2\right) \cosh \left(\frac{\alpha _d}{2}\right) \sinh \left(\frac{\alpha _r}{2}\right)\right)}{\alpha _d^2 \left(\alpha _r^2+\pi ^2\right) \cosh
   \left(\frac{\alpha _d}{2}\right) \sinh \left(\frac{\alpha _r}{2}\right)+\left(\alpha _d^2+\pi ^2\right) \alpha _r^2 \sinh \left(\frac{\alpha _d}{2}\right) \cosh
   \left(\frac{\alpha _r}{2}\right)}\; ,
\end{equation}
\end{widetext}

with $\alpha_r=T_0/\left(2\tau_r\right)$ and $\alpha_d=T_0/\left(2\tau_d\right)$. Note that if $\tau_r=\tau_d=\tau$ we have the expected simple result $\tan\left(\phi\right)=- 2\pi f \tau$.

 $\Delta x_{\rm{av}}$, the time-averaged value of the modulated fraction $\Delta x(t)$, is obtained from the experimental signals as follows:

\begin{eqnarray}
    \frac{V^{\rm{p}}_{\rm{f}}}{V^{AF}_0}&=&\frac{1}{\sqrt{2}}\frac{(-R_{AF}+R_{FM})}{R_{AF}}\Delta x^p_f\\
    \frac{V^{\rm{q}}_{\rm{f}}}{V^{\rm{p}}_{\rm{f}}}&=&\tan(\phi)=-2 \pi f \tau\\
    \Delta x_{\rm{av}}&=&\frac{1}{2}\frac{\pi}{2 \cos(\phi)^2}\Delta x^p_f\; ,
\end{eqnarray}
where $\Delta x^p_f$ is the in-phase component of $\Delta x_f$ and $V^{AF}_0$ is the DC reflectance signal in the AF phase. The $1/\sqrt{2}$ factor accounts for the lock-in detection of the rms value of the signal. Hence we have:

\begin{equation}
    \Delta x_{\rm{av}}=\frac{\pi \left(1+\left(V^{\rm{q}}_{\rm{f}}/V^{\rm{p}}_{\rm{f}}\right)^2\right)}{2 \sqrt{2}}\frac{R_{AF}}{(-R_{AF}+R_{FM})}\frac{V^{\rm{p}}_{\rm{f}}}{V^{AF}_0}\; .
\end{equation}

\section{\label{sec:Nonlinearities}Non-linearities}

\begin{figure}[htb]
\centering
\includegraphics[width=0.99\columnwidth]{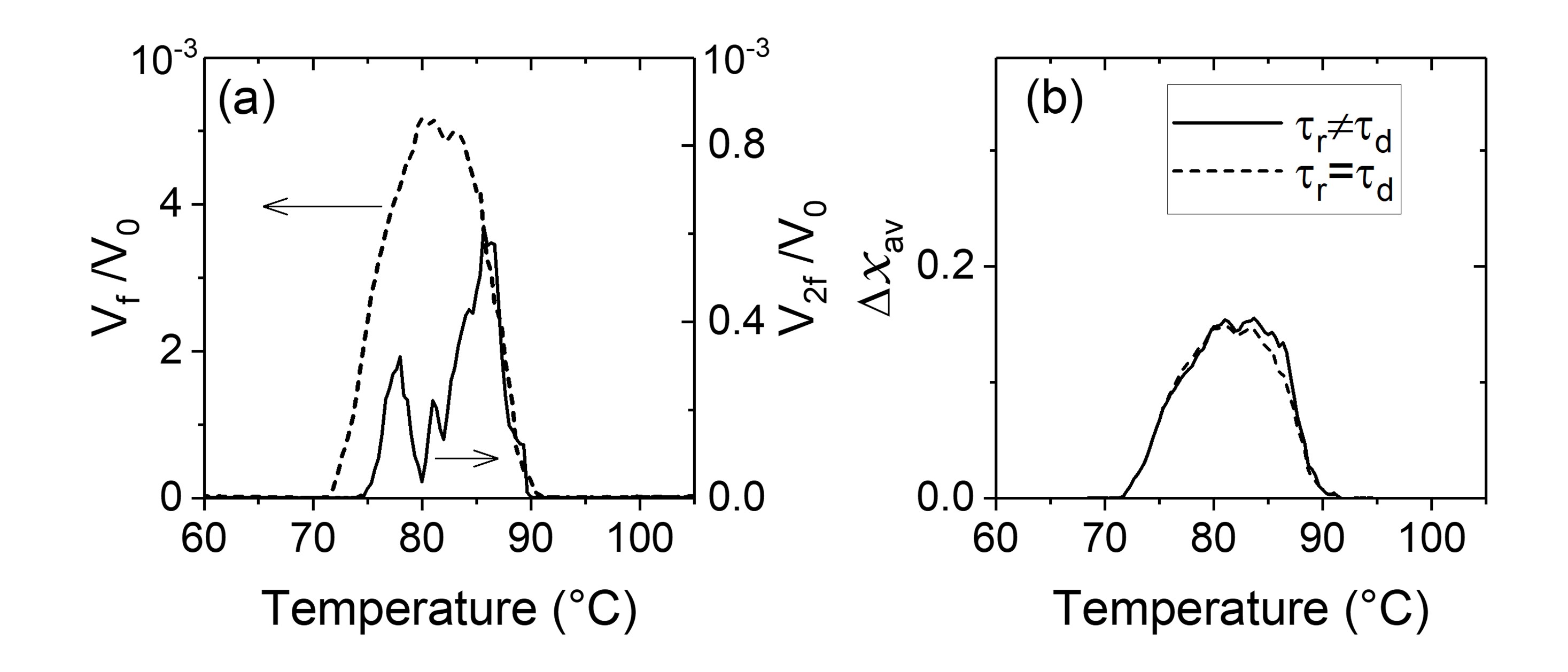}
\caption{\label{fig:f2f}  (a)  Modulated reflectance amplitudes at frequencies $f$=100\thinspace kHz and $2f$ at pump power $\approx$6.4 mW (average of the warming and cooling temperature scans). (b) Average modulated FM fraction obtained under the assumption of equal rise and decay times of the modulated reflectance,  $\tau_r$ and $\tau_d$, respectively (dashed black line), or different $\tau_r$ and $\tau_d$ (solid black line) obtained from the $2f$ and $f$ amplitudes (see text). }
\end{figure}

The non-linear behavior of the modulated reflectance signal in the mixed phase also appears as a signal at 2$f$, \textit{i.e.} twice the modulation frequency, which is absent in the square-modulated pump laser. Figure \ref{fig:f2f}\thinspace (a) show the $f$ and $2f$ signals, the latter being about 10 times smaller than the former. Within our approximation of exponential rise and decay of the signal during the ON and OFF half-period of the pump laser, this 2f-signal would result from different rise and decay times $\tau_r$ and $\tau_d$, respectively. Indeed, this is consistent with oscilloscope time traces whose detailed analysis is beyond the scope of this paper and will be the subject of a forthcoming publication. Using the calculations of Appendix \ref{App:Deltaxav} and similar formula for the $2f$-voltage we obtain $\tau_r(T)=T_0/\left(2\alpha_r\right)$ and  $\tau_d(T)=T_0/\left(2\alpha_d\right)$ from two sets of experimental data, namely the phase $\phi_{exp}$ of the $V_{\rm{f}}$ signal and the ratio $r_{exp}=\frac{V_{\rm{2f}}}{V_{\rm{f}}}$ of the $V_{\rm{2f}}$ and $V_{\rm{f}}$ amplitudes (defined above in Sec.\ref{sec:Setup}) by solving a set of two equations for each temperature.
\begin{eqnarray}
    \left(\tan\phi_{\rm{f}}\left(\alpha _r,\alpha _d\right)\right)_{\rm{calc}}&=&\tan\phi_{exp}\\
    \left(\frac{V_{\rm{2f}}}{V_{\rm{f}}}\left(\alpha _r,\alpha _d\right)\right)_{\rm{calc}}&=&r_{exp}\, ,
\end{eqnarray}
where "$\rm{calc}$" indicates the calculated quantities. $\Delta x_{\rm{av}}(T)$ is then obtained using these two times (full black line in Fig.~\ref{fig:f2f} (b)). It is seen in that this curve differs very little from the one calculated using $\tau_r(T)=\tau_d(T)$ and only data at frequency $f$  (dashed black line). This justifies the use of only $f$-frequency data  to obtain the  $\Delta x_{\rm{av}}(T)$ curves for different pump powers in Fig.~\ref{fig:XandDeltaX}.

\section{\label{App:Map}Map of transition temperatures}
\begin{figure}[htb]
\centering
\includegraphics[width=0.99\columnwidth]{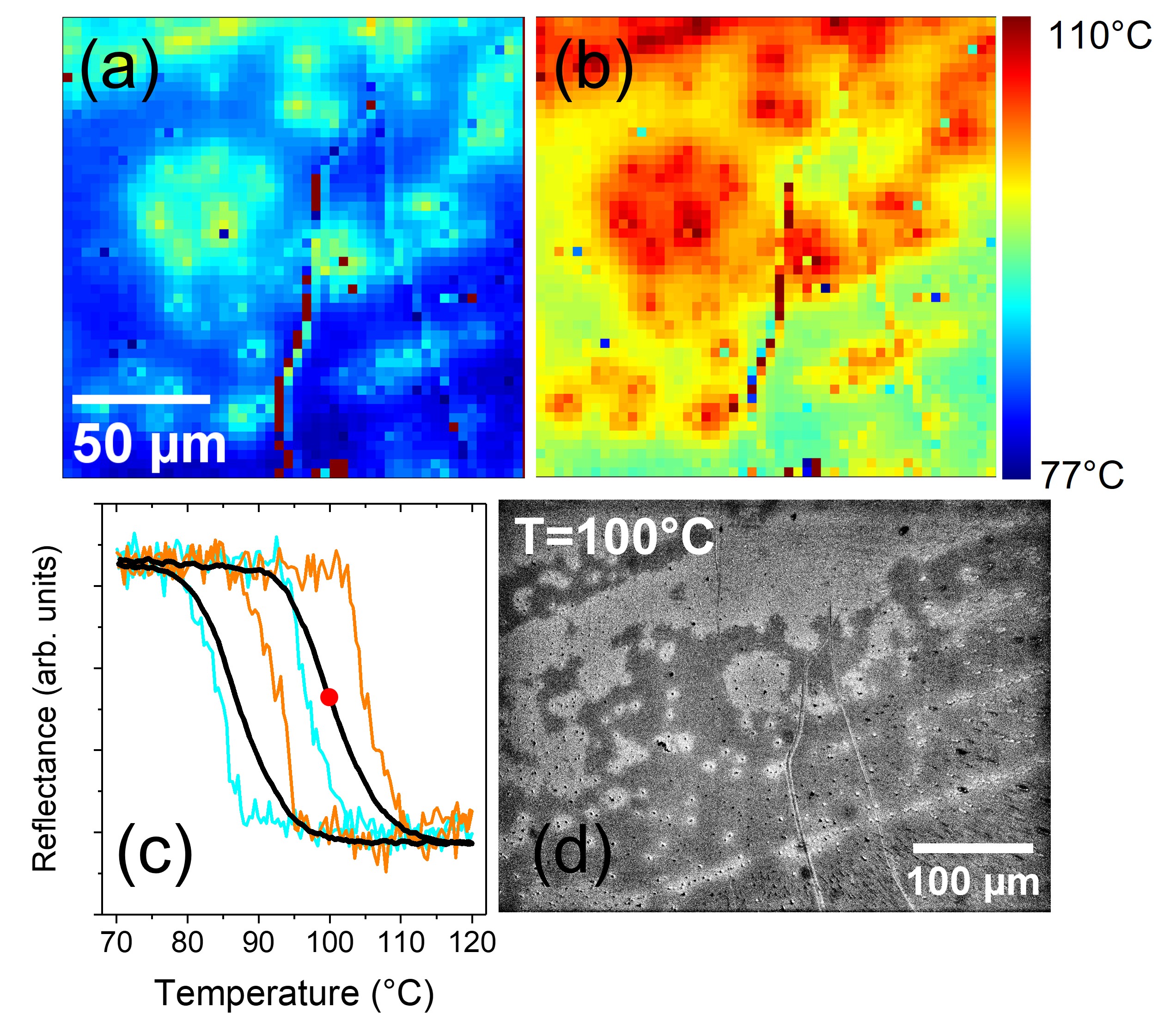}
\caption{\label{fig:Map} Map of (a) the FM→AF transition temperature and (b)
the AF→FM transition temperature. These maps show the spatial
inhomogeneity of the transition temperatures and the correlation between
both temperatures. (c) The
black curve is the reflectance averaged over a 460×343\thinspace $\upmu$m$^2$ area,
the cyan and orange curves are two local reflectance hysteresis cycles
averaged over a 3.3×3.3\thinspace $\upmu$m$^2$ area showing the temperature shift of the hysteresis cycle 
but the conservation of its width.  (d) Reflectance image at T=100$^\circ$C on the heating curve (red dot in (c)) showing AF domains with high reflectance (light gray) and FM domains with low reflectance (dark gray). (a) and (b) maps are obtained from the center region of image (d).}
\end{figure}

Figure \ref{fig:Map}(a,b) shows maps of the FM-to-AF transition temperature upon cooling, and AF-to-FM transition temperature upon heating, respectively. The hysteresis cycles are obtained by monitoring the reflectance in a microscopy setup. The transition temperatures are taken at the mid-height of a local hysteresis cycle averaged on a 3.3$\times$3.3\thinspace $\upmu$m$^2$ as shown in Fig.\thinspace \ref{fig:Map}(c). As shown in (c), local cycles exhibit similar shapes, characterized by an initial steep part followed by a smoother one till reaching the single phase region. A distribution of local loops showing slightly different transition temperature, and loop amplitude is observed. Two typical examples are shown in Fig.~\ref{fig:Map}(c) (orange, and cyan cycles) against the loop resulting from averaging over all the local cycles (black line).

\section{Model: identification and Implementation\label{App:Hysteron-Model}}

Thermal hysteresis probed on a $4~\upmu$m scale makes visible the discontinuous nature of the transition kinetics. At this scale the evolution of the heating, and cooling curves as a sequence of jumps to be associated with elementary instabilities becomes apparent, as shown in Fig \ref{fig:Fig1}, without needing a detailed analysis of the signal noise. In addition, the shape of the measured minor loops, and of the first-order reversal curves shown in Fig.~\ref{fig:Fig1}, are signatures of a Preisach distribution of the form $p(g_c, g_u) = p_1 (g_c) p_2(g_u)$ where $p_1$, and $p_2$ are expected to be rather narrow distributions of the energy barrier, and of the asymmetry between the two energy minima defining the elementary bistable unit \cite[Chapters $13 - 14$]{Bertotti1998}. This factorization is the one expected when hysteresis is mainly due to nucleation processes, and nucleation has been reported to be the key process underpinning thermal hysteresis in FeRh films \cite{Keavney2018-1, Arregi2023-1}. 

Due to the easily detectable macroscopic jumps characterizing the heating, and cooling major hysteresis branches, extrapolating a distribution of switching temperatures, from the AF to the FM phase, and vice versa is the easiest way towards model identification. To work out a distribution of bistable units from a distribution of elementary thermal hysteresis cycles, a guess is needed on the free energies of the pure phases. Following \cite{Basso2007}, the free energies of the AF, and FM phases, $g_{AF}$ and $g_{FM}$ respectively, are estimated through the following expressions:

\begin{figure}[ht]
\includegraphics[width=0.8\columnwidth]{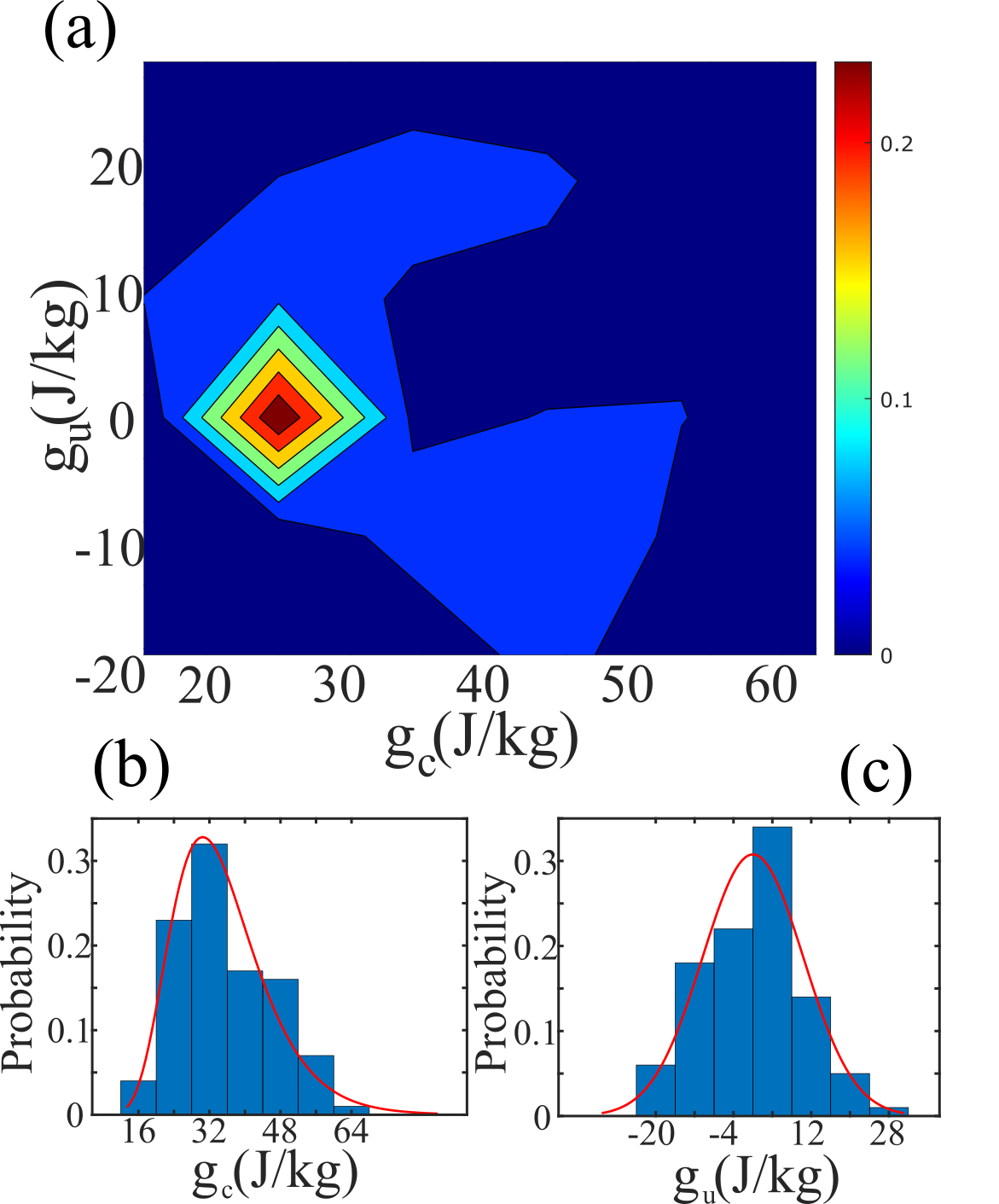}
\caption{(a) bistable distribution represented over the $(g_c, g_u)$ plane with colors representing the magnitude of the $p(g_c, g_u) = p_1(g_c) p_2(g_u)$ distribution. (b) $p_1(g_c)$ distribution with a log-normal fit (red line); (c) $p_2(g_u)$ function with a Gaussian fit (red line).}
\label{fig:preisach-distribution}
\end{figure}

\begin{eqnarray}
    g_{AF} = f_{AF}(T_0)+\left. \frac{df_{AF}}{dT} \right|_{T_0}(T-T_0) \label{eq:g_AF}, \\
        g_{FM} = f_{FM}(T_0)+\left.\frac{df_{FM}}{dT}\right|_{T_0}(T-T_0) , \label{eq:g_FM}
\end{eqnarray}
    
Hence $Z$ can be written as:

\begin{equation}\label{eq:Z-f}
    Z = \frac{1}{2} \Delta f(T_0) +\frac{1}{2} \left. \frac{d \Delta f}{dT} \right|_{T_0} (T - T_0) - \frac{1}{2} \gamma\;  x,
\end{equation}

where $\Delta f = f_{FM} - f_{AF}$, and the second term on the right is the latent heat associated to the phase transition with $d \Delta f / dT = \Delta s$ its entropy. The last term in (\ref{eq:Z-f}) is an additional mean field contribution taking into account the average magnetic, and elastic interactions between units (see \cite{Della-Torre1966}, and Chapter~$2$ in \cite{Mayergoyz2003} for a general overview on the use of a mean-field term within the Preisach model). The value $\Delta s = 12.5$~Jkg$^{-1}$K$^{-1}$ has been taken from the literature \cite{Lyubina_2017} while the mean field parameter has been adjusted on the data to $\gamma =  3$~J\thinspace kg$^{-1}$. Taking $T_0$ as the temperature where the free energy of the two pure phases are equal, namely $f_{AF}(T_0) = f_{FM}(T_0)$, we have $\Delta f(T_0) = 0$.
In this way, following the switching rules (\ref{eq:switching}), the relationship between the $g_c^{(i)}$, and $g_u^{(i)}$ parameters defining the $i$-th elementary unit, and the temperatures $\alpha^{(i)}$ where the unit switches from AF to FM when heating, and $\beta^{(i)}$ where the FM to AF transition takes place on cooling, can be written as follows:

\begin{align}
    &g_u^{(i)} = \frac{1}{2} \left[ \Delta f(T_0) + \left.\frac{d\Delta f}{dT}\right|_{T_0} \left( \frac{\alpha^{(i)}+\beta^{(i)}}{2} - T_0 \right) - \gamma \; x \right] \label{eq:g_u-alpha-beta}\\
    &g_c^{(i)} = \frac{1}{2} \left.\frac{d\Delta f}{dT}\right|_{T_0} \frac{\alpha^{(i)}-\beta^{(i)}}{2}. \label{eq:g_c-alpha-beta}
\end{align}

The $\alpha^{(i)}$, and $\beta^{(i)}$ distributions have been identified as follows. The  \textit{findchangepts} \uppercase{MATLAB} function has been used on the major heating, and cooling curves to extract the temperatures where abrupt jumps have been measured. The algorithm used by Matlab is detailed in  \cite{Killick-2012}.

Afterwards, the overall phase change (\textit{i.e.} reflectivity difference) between the two pure phases has been discretized as the sum over $10^2$ elementary instabilities. This allows to associate to each detected jump a number of elementary units proportional to the ratio between its amplitude and the total change between the pure phases. 
Imposing as a constraint $\alpha^{(i)} > \beta^{(i)}$, and considering $\alpha^{(i)}$ and $\beta^{(i)}$ as independent random variables the distributions of $g_c^{(i)}(\alpha^{(i)} - \beta^{(i)})$ and $g_u^{(i)}( \alpha^{(i)} + \beta^{(i)})$ have been extracted. The result is shown in Fig.~\ref{fig:preisach-distribution}. 
Because of the discrete nature of the bistable unit ensemble, the FM phase fraction is computed through a sum over the units state, 
\begin{equation}
    x = \sum_j \; \sum_{i}  p(g_c^{(j)}, g_u^{(i)}) \qquad \mbox{with} \qquad  g_u^{(i)} \leq b(g_c^{(i)})
\end{equation}

Distributions $p_1(g_c)$ and $p_2(g_u)$ in Figure \ref{fig:preisach-distribution} can have been fitted with a lognormal and a Gaussian respectively.

\end{document}